\documentclass[a4paper,twocolumn,floatfix,10pt,superscriptaddress]{quantumarticle}
\pdfoutput=1

\usepackage[utf8]{inputenc}

\usepackage[T1]{fontenc}
\usepackage{amsmath}
\usepackage{physics}
\usepackage{siunitx} 
\usepackage{tikz}
\usepackage{lipsum}

\usepackage[acronym]{glossaries}

\usepackage[numbers,sort&compress]{natbib}
\usepackage[colorlinks=true,pdfencoding=auto]{hyperref}
\usepackage[capitalise]{cleveref}
\usepackage[table]{xcolor}

\usepackage{graphicx}
\usepackage{subfigure}
\usepackage{amssymb}
\usepackage{multirow}
\usepackage{soul}

\glsdisablehyper
\newacronym{HBARs}{HBARs}{High-Overtone Bulk Acoustic-wave Resonators}
\newacronym{PA}{PA}{parametric amplifier}
\newacronym{TLS}{TLS}{two-level system}
\newacronym{BS}{BS}{beam-splitter}
\newacronym{CPS}{CPS}{correlated photon source}
\newacronym{AE}{AE}{adiabatic elimination}
\newacronym{ME}{ME}{master equation}
\newacronym{QRT}{QRT}{quantum regression theorem}
\newacronym{NZ}{NZ}{Nakajima-Zwanzig}
\newacronym{CPTP}{CPTP}{completely positive and trace preserving}
\newacronym{TP}{TP}{trace preserving}
\newacronym{CP}{CP}{completely positive}
\newacronym{CV}{CV}{continuous variable}
\newacronym{TMS}{TMS}{two-mode squeezed}
\newacronym{RWA}{RWA}{rotating wave approximation}
\newacronym{SW}{SW}{Schrieffer-Wolff}
\newacronym{ED}{ED}{exact-diagonalization}
\newacronym{QMS}{QMS}{qubit-mediated squeezing}
\newacronym{RP}{RP}{resolved Purcell}

\newcommand{\etal}{et al.}

\newcommand{\ddt}{\frac{\mathrm{d}}{\mathrm{d}t}}

\usepackage{xcolor}
\definecolor{aprcolor}{rgb}{0.9,0.1,0.9}

\begin{document}
\title{A Minimal Photon Source for Remote Entanglement Stabilization}

\title{Minimal Photon Source for Optimized Distributed Entanglement Stabilization}

\title{Driven two-level systems as a minimal resource for remote/distributed entanglement stabilization}

\title{Driven two-level systems as a minimal resource for remote entanglement stabilization}

\author{Philippe Gigon}
\email{philippe.gigon@protonmail.com}
\orcid{0009-0008-9802-9261}
\affiliation{Technical University of Munich, TUM School of Natural Sciences, Physics Department, 85748 Garching, Germany}
\affiliation{Walther-Meißner-Institut, Bayerische Akademie der Wissenschaften, 85748 Garching, Germany}
\affiliation{Munich Center for Quantum Science and Technology (MCQST), 80799 Munich, Germany}
\author{Adrian Parra-Rodriguez}
\orcid{0000-0002-0896-9452}
\affiliation{Technical University of Munich, TUM School of Natural Sciences, Physics Department, 85748 Garching, Germany}
\affiliation{Walther-Meißner-Institut, Bayerische Akademie der Wissenschaften, 85748 Garching, Germany}
\affiliation{Munich Center for Quantum Science and Technology (MCQST), 80799 Munich, Germany}
\author{Joan Agustí}
\orcid{0000-0002-9883-1958}
\affiliation{Institute of Fundamental Physics IFF-CSIC, Calle Serrano 113b, 28006 Madrid, Spain}
\author{Peter Rabl}
\orcid{0000-0002-2560-8835}
\affiliation{Technical University of Munich, TUM School of Natural Sciences, Physics Department, 85748 Garching, Germany}
\affiliation{Walther-Meißner-Institut, Bayerische Akademie der Wissenschaften, 85748 Garching, Germany}
\affiliation{Munich Center for Quantum Science and Technology (MCQST), 80799 Munich, Germany}

\begin{abstract}
We analyze the autonomous stabilization of remote entanglement by driving two distant qubits with the output of a correlated photon source. By treating the qubits as idealized entanglement detectors, we develop a general framework to quantify the maximum amount of entanglement that can be remotely stabilized in this way with a given photon source. We then apply this approach to evaluate the suitability of a single driven two-level system  as a minimal resource for autonomous entanglement distribution schemes. While our analysis confirms the presence of distributable entanglement in the Mollow sidebands of a bare two-level system, we show that stabilizing close to maximally entangled states requires additional filter cavities that enhance the relevant correlated emission events compared to other processes. We identify optimized driving and cavity parameters and explain the achievable amount of entanglement in different regimes in terms of an effective two-mode squeezing model. These insights are particularly relevant for quantum networks based on photons or phonons in solid-state systems, where isolated spins, impurity centers, or other two-level defects are readily available, while alternative sources of correlated photons are difficult to realize.  
\end{abstract}

\maketitle

\section{Introduction}
Distributing entanglement across different nodes of a quantum network is a vital requirement for many quantum communication~\cite{gisin_quantum_2007,kimble_quantum_2008} and distributed quantum computing~\cite{jiang_distributed_2007, monroe_large-scale_2014, gottesman_demonstrating_1999, cacciapuoti_quantum_2020, beukers_remote-entanglement_2024, awschalom_development_2021} applications. Most conventional schemes for realizing this task rely on quantum channels that directly connect the nodes, over which quantum correlations can be established, for example, via the emission and reabsorption of photons or phonons~\cite{cirac_quantum_1997,stannigel_driven-dissipative_2012,ritter_elementary_2012,northup_quantum_2014,pichler_quantum_2015,reiserer_cavity-based_2015,axline_on-demand_2018,kurpiers_deterministic_2018,bienfait_phonon-mediated_2019,kurpiers_quantum_2019,leung_deterministic_2019,zhong_deterministic_2021,lingenfelter_exact_2024,shah_stabilizing_2024,almanakly_deterministic_2025,irfan_autonomous_2025}.
Alternatively, entanglement can be established over long distances by coupling two or more qubits to a common distributed reservoir with preexisting quantum correlations~\cite{son_entanglement_2002,benatti_environment_2003,kraus_discrete_2004,paternostro_complete_2004,felicetti_dynamical_2014,didier_remote_2018,you_waveguide_2018,ma_stabilizing_2019,zippilli_entanglement_2013,pocklington_universal_2024,govia_stabilizing_2022,agusti_long-distance_2022,agusti_autonomous_2023,andres-juanes_entangling_2025}. This approach has the advantage that it eliminates the need for directly connecting all the individual nodes of the network, it can be used to entangle systems with very dissimilar frequencies, and the preparation of the correlated photonic bath can be implemented and optimized independently of the local network nodes. Moreover, entanglement is distributed fully autonomously through a passive coupling to a reservoir, requiring no measurements or feedback control.

In their seminal work~\cite{kraus_discrete_2004}, Kraus and Cirac first predicted that two separated qubits can be driven into a pure, maximally entangled state by coupling them to a shared two-mode squeezed (TMS) photonic reservoir. Building on this initial idea, various protocols have been proposed to engineer and generalize this scheme in extended waveguide QED systems~\cite{didier_remote_2018,govia_stabilizing_2022,agusti_long-distance_2022,agusti_autonomous_2023} and coupled cavity arrays~\cite{zippilli_entanglement_2013}. A general approach for creating a distributed TMS reservoir is to couple the output of a non-degenerate parametric amplifier to two separate waveguides. When the bandwidth of the amplifier is sufficiently broad~\cite{agusti_long-distance_2022}, this setting fulfills the requirements established in~\cite{kraus_discrete_2004} and can be used, in principle, to distribute pure and close-to-maximally entangled qubit states. While parametric amplifiers are well-developed in the optical and microwave regimes, where this entanglement distribution scheme was recently experimentally demonstrated~\cite{andres-juanes_entangling_2025}, they may not be readily available in other physical platforms or may be difficult to integrate without substantial experimental overhead. It is therefore of interest to search for potential alternative sources of correlated photons suited for remote entanglement distribution, but that may be simpler to realize or more commonly available.

In this work, we revisit the autonomous distribution of entanglement between two remote qubits and address the suitability of a generic \gls{CPS} for this task. For this purpose, we motivate and derive a quantitative measure for the amount of \emph{distributable entanglement}, which can be computed solely from two-time correlation functions of the CPS. We then use this measure to investigate the use of a single driven \gls{TLS} as a minimal resource for entanglement distribution. While a \gls{TLS} can emit only a single photon at a time, it is well known that the emission spectrum of a strongly driven \gls{TLS} exhibits a characteristic three-peak structure, the Mollow triplet~\cite{mollow_power_1969}, with photons emitted at different frequencies exhibiting strong correlations.
These correlations arise from rich multi-photon processes~\cite{carmichael_quantum-mechanical_1976,cohen-tannoudji_dressed-atom_1977,kimble_photon_1977,dagenais_investigation_1978,aspect_time_1980,dalibard_correlation_1983,arnoldus_photon_1984} and have been systematically analyzed in many theoretical~\cite{del_valle_theory_2012,valle_distilling_2013,gonzalez-tudela_two-photon_2013,sanchez_munoz_violation_2014,munoz_emitters_2014,chang_deterministic_2016,lopez_carreno_photon_2017,sanchez_munoz_filtering_2018,zubizarreta_casalengua_two-photon_2023,lopez_carreno_entanglement_2024,bermudez-feijoo_spectral_2025,elliott_quantum-correlated_2025,vivas-viana_nonclassical_2025} and experimental~\cite{peiris_two-color_2015,peiris_franson_2017,hanschke_origin_2020,masters_simultaneous_2023,liu_violation_2024,wang_purcell-enhanced_2025,yang_entanglement_2025} works. Importantly, it was shown that the photons emitted from a driven \gls{TLS} can exhibit non-classical correlations~\cite{sanchez_munoz_violation_2014} and become entangled~\cite{lopez_carreno_entanglement_2024,vivas-viana_nonclassical_2025}, as observed in experiments with optical emitters~\cite{peiris_franson_2017,liu_violation_2024,wang_purcell-enhanced_2025} and driven superconducting qubits~\cite{yang_entanglement_2025}. 
Using these quantum correlations, L\'opez Carre\~no \textit{et al.}~\cite{lopez_carreno_entanglement_2024} proposed and analyzed a scheme to entangle two distant bosonic nodes with photons scattered from a driven \gls{TLS}.
The total amount of entanglement generated by the \gls{TLS}, however, has been found to be rather low~\cite{lopez_carreno_entanglement_2024,yang_entanglement_2025,vivas-viana_nonclassical_2025}, unless the photonic states are postselected~\cite{lopez_carreno_entanglement_2024,peiris_franson_2017,liu_violation_2024,wang_purcell-enhanced_2025}.

Our current approach quantifies these general observations in terms of the distributable entanglement and provides additional intuition by associating the relevant photonic correlations with an equivalent \gls{TMS} state. This identification reveals stringent limits on both the effective squeezing strength and the purity of the generated photons. To overcome these limitations, we propose to couple the \gls{TLS} to additional filter cavities to structure the effective density of states and thereby enhance and purify the photonic correlations. We investigate this setup under various driving and coupling conditions, using both exact numerical simulations and approximate analytical models, and evaluate the distributable entanglement across different parameter regimes. This analysis demonstrates how this extended CPS, i.e., a driven \gls{TLS} in a structured environment, can be systematically optimized to approach the performance of an ideal TMS source. Using an approximate analytical model for the combined TLS-cavity system in the sideband-resolved regime, we further identify a common underlying mechanism that gives rise to optimal squeezing correlations under both weak- and strong-coupling conditions.  

The remainder of this article is structured as follows. In~\cref{Sec:Model}, we start by introducing the general setting of a \gls{CPS} emitting into two waveguides, from which we motivate and derive the distributable entanglement as a quantitative measure to characterize an arbitrary \gls{CPS}.  
In~\cref{Sec:DrivenTLS}, we apply this general formalism first to the case of a bare \gls{TLS} emitting into two broadband waveguides, after which we treat the extended setting with additional filter cavities in~\cref{Sec:MollowCavities}. Based on these results, in~\cref{Sec:DressedStateAnalysis}, we perform a more detailed discussion of the general mechanism that underlies the generation of strongly-correlated photonic fields in both the weak- and strong-coupling regimes. Finally, in~\cref{Sec:Conclusion}, we summarize the main conclusion and briefly outline potential experimental platforms where these findings can be tested and applied. 

\section{Autonomous entanglement distribution}\label{Sec:Model}

\begin{figure}[t]
    \centering
    \includegraphics[width=0.9\linewidth]{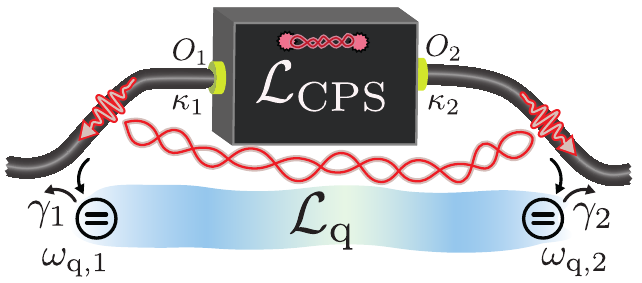}
        \caption{Schematics of the basic chiral quantum network considered in this work. Here, a source (black box) produces a stream of correlated photons that drive two distant target qubits into an entangled steady state. See text for more details.}
    \label{fig:Auto_entang_dist}
\end{figure}

For the following analysis, we consider a setup as depicted in~\cref{fig:Auto_entang_dist}. Here, the black box represents a generic source of correlated photons, driven by an external classical field and having two output ports, $i=1,2$. At these ports, photons are emitted into two separate waveguides with rates $\kappa_i$ and drive two distant qubits (which we refer to as \textit{target} qubits) located further along the channels. The qubits have transition frequencies $\omega_{{\rm q},i}$ and decay into the broadband waveguide with rates $\gamma_i$. For conceptual simplicity, we assume a fully unidirectional qubit-waveguide coupling~\cite{carmichael_quantum_1993,gardiner_driving_1993,lodahl_chiral_2017,soro_chiral_2022,suarez-forero_chiral_2025,barzanjeh_nonreciprocity_2025} to avoid any backaction of the qubits on the photon source, and we also assume that all parts of the network are sufficiently cold such that thermal excitations can be neglected.

\subsection{Model}
Under these assumptions, the dynamics of the combined state $\rho$ of the photon source and the target qubits can be described by a cascaded \gls{ME} of the form ~\cite{carmichael_quantum_1993,gardiner_driving_1993,gardiner_driving_1994,pichler_quantum_2015}
\begin{align}
\dot \rho(t)=\left[\mathcal{L}_{\mathrm{CPS}}(t)+\mathcal{L}_{\rm Q}+\mathcal{L}_{\rm I}\right]\rho(t).
    \label{Eq:FullSyst}
\end{align}
Here, $\mathcal{L}_{\mathrm{CPS}}(t)$ and $\mathcal{L}_{\rm Q}$ determine the individual dynamics of the driven photon source and the qubits, respectively, and $\mathcal{L}_{\rm I}$ accounts for their waveguide-mediated interaction. For the current cascaded setting, we obtain
 \begin{align}    
 \mathcal{L}_{\rm I}=\sum_i\sqrt{\kappa_{i} \gamma_{i}} \Big( [O_i \bullet, \sigma_i^{+}] + [\sigma_i^{-}, \bullet O_i^\dag] \Big),
     \label{Eq:CascadedInteraction}
 \end{align}
where we have neglected retardation effects and already absorbed any propagation phases into an appropriate redefinition of the qubit states~\cite{agusti_long-distance_2022}. 

In the absence of the \gls{CPS}, the qubits are assumed to be noninteracting and only coupled to the respective waveguides. Their undriven evolution is thus described by the Liouville operator
 \begin{align}
     \mathcal{L}_{\rm Q}=
     -i\sum_i\frac{\omega_{\mathrm{q},i}}{2} \left[ \sigma_i^z,\bullet\right]+ \sum_i\gamma_{i} \mathcal{D}[\sigma_i^-],
 \end{align}
where $\mathcal{D}[A]\equiv A\bullet A^\dag-\{A^\dag A,\bullet\}_+/2$. The dynamics of the photon source is determined by the Liouville operator 
\begin{align}
    \mathcal{L}_{\mathrm{CPS}}(t)=\mathcal{L}^0_{\mathrm{CPS}}(t)+\sum_i\kappa_{i}\mathcal{D}[O_i],
    \label{Eq:CPSLindbladian}
\end{align}
where the jump operators $O_i$ determine the coupling to the waveguides. The bare dynamics described by $\mathcal{L}^0_{\mathrm{CPS}}(t)$ depends on the specifics of the sources and is, for now, left unspecified.

We remark that the \gls{ME} given in Eq.~\eqref{Eq:FullSyst} still contains an explicit time-dependence from the external field that is driving the photon source. In all the examples considered below, the photon source is only driven by a single monochromatic field of frequency $\omega_{\rm d} \approx \omega_{\mathrm{q},i}$. In this case, this time dependence can be removed by the unitary transformation,  
\begin{equation}
O_i \rightarrow O_i e^{-i\omega_{\rm d} t}, \qquad \sigma_i^-\rightarrow \sigma_i^- e^{-i\omega_{\rm d} t},
\label{Eq:RotFrame}
\end{equation}  
and by applying a rotating-wave approximation (RWA) to neglect fast oscillating terms $\sim e^{\pm 2i \omega_{\rm d} t}$. Other scenarios might involve different driving schemes and resonance conditions, but in general, similar rotating frames, in which the ME becomes time-independent, can be identified. 

\subsection{A correlated photonic reservoir} \label{Sec:AEBB}

In the setup introduced above, correlated photons emitted from the source continuously excite the two qubits and can drive them into an entangled stationary state. In previous studies~\cite{didier_remote_2018,agusti_long-distance_2022,agusti_autonomous_2023,andres-juanes_entangling_2025}, this process has already been extensively analyzed for the case of a non-degenerate parametric amplifier, which produces a Gaussian \gls{TMS} state at the outputs. In this case, the steady state of the qubits can approach a maximally entangled Bell state in the limit of large squeezing and when the bandwidth of photons exceeds the qubit decay, $\kappa \gg \gamma$~\cite{agusti_long-distance_2022}.  

Here, we are interested in the suitability of a generic source of correlated photons to achieve the same task. To do so, we consider two ideal target qubits in the limit $\gamma_i\rightarrow 0$, i.e., under conditions where their dynamics are slow compared to all relevant timescales of the \gls{CPS}. In this limit, we evaluate the reduced steady state of the qubits, $\rho^{\rm ss}_{\mathrm{q}}={\rm Tr}_\mathrm{CPS}\{\rho(t\rightarrow \infty)\}$, and use the concurrence~\cite{horodecki_quantum_2009} of this state, $\mathcal{C}(\rho^{\rm ss}_\mathrm{q})\equiv \mathcal{C}_\mathrm{d}$, as a quantitative measure for the amount of qubit-qubit entanglement that can be remotely stabilized by the source. In other words, the qubits serve as detectors for the amount of entanglement that is emitted by the source in a narrow frequency band around the qubit frequencies $\omega_{\mathrm{q},1}$ and $\omega_{\mathrm{q},2}$. The benefit of this operational definition  is that it becomes independent of any details of the target qubits, and different CPSs can be directly compared without analyzing the full quantum network.

\subsubsection{Effective ME for the target qubits}

To evaluate $\mathcal{C}_\mathrm{d}$, we make use of the assumed separation of timescales, $\gamma_i\ll {\rm min}\{\kappa_1,\kappa_2\}$, and eliminate the fast degrees of freedom of the photon source in the standard Born-Markov approximation~\cite{breuer_theory_2002,rivas_open_2012}. As a result 
we obtain a \gls{ME} for the reduced state of the qubits, $\dot{\rho}_\mathrm{q}(t)=\mathcal{L}_\mathrm{q} \rho_\mathrm{q}(t)$. In the rotating frame of the drive defined in~\cref{Eq:RotFrame} above, this \gls{ME} is of the general form (see~\cref{Sec:AdiabaticEliminationCPS} for more details)
\begin{align}
&\mathcal{L}_{\mathrm{q}}=\mathcal{L}_{\rm Q}+\sum_i \sqrt{\gamma_{i}}\left[(\varepsilon_i^* \sigma_i^--\varepsilon_i\sigma_i^+),\bullet \right] \label{Eq:CPSME}\\
&+\sum_{s_{1,2} =\pm}\sum_{i,j}\sqrt{\gamma_{i}\gamma_{j}}C^{ij}_{s_1s_2}(\Delta_{\mathrm{q},j}) s_1s_2\left[\left[\bullet,\sigma_j^{-s_1} \right] ,\sigma_i^{-s_2}\right].
\nonumber
\end{align}
Here, $\Delta_{\mathrm{q},i}= \omega_{\mathrm{q},i}-\omega_{\mathrm{d}}$ are the detunings of the qubits from the drive, and the $C^{ij}_{s_1s_2}$ are the Laplace-transformed two-time correlation functions of the jump operators,
\begin{align}
\begin{aligned}
    C^{ij}_{--}(\Delta_{\mathrm{q},j})&=\sqrt{\kappa_i\kappa_j}\int_0^\infty \mathrm{d}\tau e^{-i\Delta_{\mathrm{q},j} \tau}\langle \overline{O}_i(\tau) \overline{O}_j\rangle_{\mathrm{ss}}, \\
    C^{ij}_{-+}(\Delta_{\mathrm{q},j})&=\sqrt{\kappa_i\kappa_j}\int_0^\infty \mathrm{d}\tau e^{-i\Delta_{\mathrm{q},j} \tau}\langle \overline{O}_i^\dag(\tau) \overline{O}_j\rangle_{\mathrm{ss}},
    \label{Eq:BBCoefs}
\end{aligned}
\end{align}
where $\overline{O}_i=O_i-\langle O_i \rangle_{\mathrm{ss}}$ and $\langle \bullet \rangle_{\mathrm{ss}}$ denotes the steady-state expectation value of the CPS. In Eq.~\eqref{Eq:CPSME}, $\varepsilon_i=\sqrt{\kappa_{i}}\langle O_i \rangle_{\mathrm{ss}}$ represents the coherent parts of the fields emitted by the source. The remaining two coefficients occurring in~\cref{Eq:CPSME} are defined as 
$C^{ij}_{++}(\Delta_{\mathrm{q},j})\equiv [C^{ij}_{--}(\Delta_{\mathrm{q},j})]^*$
and $C^{ij}_{+-}(\Delta_{\mathrm{q},j})\equiv[C^{ij}_{-+}(\Delta_{\mathrm{q},j})]^*$.

\subsubsection{Resonant contributions}
While Eq.~\eqref{Eq:CPSME} can already be used to evaluate $\rho_{\rm q}^{\rm ss}$ for a given set of correlation functions $C_{\pm\pm}$ of the CPS, additional simplifications can be made by taking the strict limit $\gamma_{1}=\gamma_2=\gamma\rightarrow 0$. In this case, only detunings satisfying $|\Delta_{\mathrm{q},1}\pm\Delta_{\mathrm{q},2}|\simeq 0$ result in resonant contributions that significantly affect the dynamics of the target qubits. In addition, for all the examples considered in this work, entanglement generation is maximized by the choice $\Delta_{\mathrm{q},1}=-\Delta_{\mathrm{q},2}\neq 0$ and by moving into a frame rotating with these nonvanishing detunings, the effective qubit ME reduces to 
\begin{align}
    \mathcal{L}_{\mathrm{q}}=&\,\sum_i \gamma (N_i+1) \mathcal{D}[\sigma_i^-]+ \gamma N_i\mathcal{D}[\sigma_i^+] \label{Eq:SqueezingLindbladian}\\
    &- \gamma\left(M\mathcal{D}[\sigma_1^+,\sigma_2^+] +M \mathcal{D}[\sigma_2^+,\sigma_1^+] +\text{H.c.}\right),\nonumber   
\end{align}
where $\mathcal{D}[A,B]=A\bullet B-\{BA,\bullet\}/2$. The remaining coefficients that appear in these equations are related to the correlation spectra from above through
\begin{align}
\begin{aligned}
    N_i&= 2 \Re{C^{ii}_{-+}(\Delta_{\mathrm{q},i})},  \\
    M&=C_{--}^{12}(\Delta_{\mathrm{q},2})+C_{--}^{21}(\Delta_{\mathrm{q},1}).
    \label{Eq:EffectiveCoefsCPS}
\end{aligned}
\end{align}

\subsection{Entanglement and squeezing correlations}

In the simplified form given in~\cref{Eq:SqueezingLindbladian}, the effective qubit ME is equivalent to a \gls{ME} for two qubits coupled to a thermal \gls{TMS} reservoir. Following the discussion in Refs.~\cite{kraus_discrete_2004,agusti_long-distance_2022}, this reservoir can be fully characterized in terms of an equivalent two-mode state 
\begin{align}
    \rho_{\mathrm{eff}} = S(\chi) \left(\rho_{\rm{th},1} \otimes \rho_{\rm th,2} \right)S^\dagger(\chi).
    \label{Eq:EnvironmentDecomposition}
\end{align}
Here, $\rho_{{\rm th},i}$ denotes a thermal state of a single bosonic mode with annihilation (creation) operator $a_i$ ($a_i^\dag$) and   
\begin{align}
    S(\chi)=\exp( \chi^* a_1 a_2-\chi a_1^\dagger a_2^\dagger)
    \label{Eq:TMSOp}
\end{align} 
with $\chi=r_{\rm eff} e^{i\Theta}$ the \gls{TMS} operator~\cite{braunstein_quantum_2005}. The connection to the parameters in~\cref{Eq:EffectiveCoefsCPS} is established by identifying $N_i\equiv\langle a_i^\dag a_i\rangle_{\rho_{\mathrm{eff}}}$ and $M\equiv\langle a_1 a_2\rangle_{\rho_{\mathrm{eff}}}$.

\begin{figure}[t]
    \centering
\includegraphics[width=\linewidth]{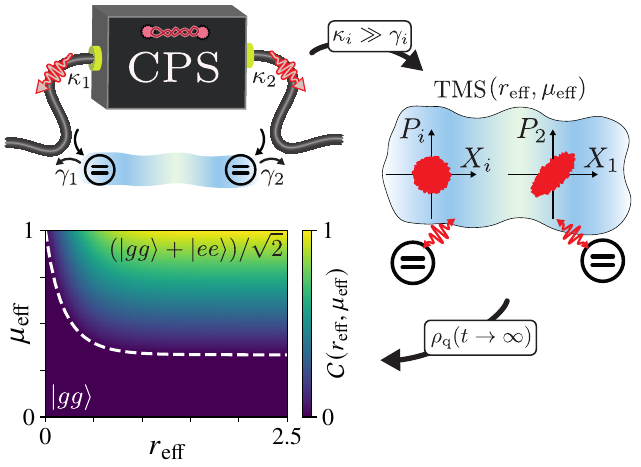}
    \caption{Operative definition of distributable entanglement.   In the Markovian regime ($\kappa_i\gg \gamma_i$), the target qubits see an effective \gls{TMS} environment, which is fully characterized by the effective squeezing strength $r_\mathrm{eff}$ and effective purity $\mu_\mathrm{eff}$. These parameters, in turn, can be evaluated from a set of two-time correlation functions of the \gls{CPS}, see~\cref{Eq:EffectiveCoefsCPS} and~\cref{Eq:TMSEff}.
    For a given $r_{\rm eff}$ and $\mu_{\rm eff}$ the concurrence $\mathcal{C}_\mathrm{d}\equiv\mathcal{C}(r_\mathrm{eff},\mu_\mathrm{eff})$ of the resulting qubit steady state is given in Eq.~\eqref{Eq:DefConc} and plotted in the lower left corner. The white dashed line marks the minimal conditions above which $\mathcal{C}_\mathrm{d}>0$. Overall, this approach allows us to characterize the \gls{CPS} in terms of a single, physically intuitive measure. See text for more details.}
    \label{fig:Misc2}
\end{figure}

Apart from the effective squeezing parameter $r_{\mathrm{eff}}$, the second quantity of interest is the purity $\mu_{\mathrm{eff}}$~\cite{serafini_entanglement_2004} of the effective \gls{TMS} state. Both quantities can be related to the coefficients in the reduced qubit ME in~\cref{Eq:SqueezingLindbladian} via 
\begin{align}\
\begin{aligned}
r_{\mathrm{eff}} &\equiv\frac{1}{2}\tanh^{-1}{\left(\frac{2 |M|}{N_1+N_2 +1}\right)}, \\
    \mu_{\mathrm{eff}}&\equiv \frac{1}{(1+2N_1)(1+2N_2)-4 |M|^2},  
    \label{Eq:TMSEff}
\end{aligned}    
\end{align}
and $M=|M| e^{i\Theta}$. In terms of $r_{\mathrm{eff}}$ and $\mu_{\mathrm{eff}}$, the concurrence of the steady state of the Lindbladian in Eq.~\eqref{Eq:SqueezingLindbladian} is given by 
\begin{align}
\begin{aligned}
    \mathcal{C}_\mathrm{d}&=\max\left\{|\mu_\mathrm{eff}\tanh(2r_\mathrm{eff})|-\frac{1-\mu_\mathrm{eff}}{2},0\right\}.
    \label{Eq:DefConc}
\end{aligned}
\end{align}
In the limit $\mu_{\text{eff}}\rightarrow 1$ and  $r_{\text{eff}}\to \infty$, the steady state of the target qubits approaches a pure, maximally entangled state~\cite{kraus_discrete_2004}
\begin{align}
    \ket{\psi}\rightarrow\frac{\ket{gg} + e^{i\Theta} \ket{ee}}{\sqrt{2}},
\end{align}
and $\mathcal{C}_\mathrm{d}\rightarrow 1$. For finite values of the effective parameters, the concurrence $\mathcal{C}_\mathrm{d}$ is plotted in~\cref{fig:Misc2}. The entanglement vanishes below the white dashed line in this plot, which corresponds to an effective purity $\mu_{\rm eff}=1/3$ in the limit of large $r_{\rm eff}$~\cite{agusti_long-distance_2022}.

\subsection{Distributable entanglement}
In summary, the derivation above shows that in the limit of ideal, infinitely long-lived target qubits, but under otherwise rather generic assumptions, the suitability of a given CPS for remote entanglement distribution can be rigorously quantified in terms of the distributable entanglement $\mathcal{C}_\mathrm{d}$. For a given pair of frequencies $\omega_{{\rm q},1}$ and $\omega_{{\rm q},2}$, this quantity depends on the parameters $N_i$ and $|M|$ only, which in turn are fully specified by the correlation spectra $C^{ij}_{\pm\pm}$ of the two output fields. Based on the relations in Eq.~\eqref{Eq:TMSEff}, these parameters can further be mapped to an effective TMS reservoir with squeezing strength $r_{\rm eff}$ and purity $\mu_{\rm eff}$. 
This mapping is particularly useful for obtaining physical insights into the properties and limitations of the CPS, which can then be discussed in terms of more familiar properties of squeezed states. Therefore, this overall approach, summarized in~\cref{fig:Misc2}, is ideally suited to compare different CPSs for autonomous entanglement distribution schemes and to identify optimized operation regimes for a given device.

\section{Entanglement from a driven TLS} \label{Sec:DrivenTLS}

Based on the general formalism developed in the previous section, we now apply this approach to quantify the amount of distributable entanglement generated by a driven \gls{TLS}. 
It has already been pointed out previously that photons emitted within certain frequency windows of a driven \gls{TLS} can exhibit nonclassical correlations~\cite{sanchez_munoz_violation_2014} and entanglement~\cite{lopez_carreno_entanglement_2024,yang_entanglement_2025,vivas-viana_nonclassical_2025,liu_violation_2024}. Here, our goal is to revisit and quantify these nonclassical correlations in light of autonomous entanglement distribution schemes, and to understand and overcome potential limitations.

\subsection{Model} 

\begin{figure}[t]
    \centering
    \includegraphics[width=\linewidth]{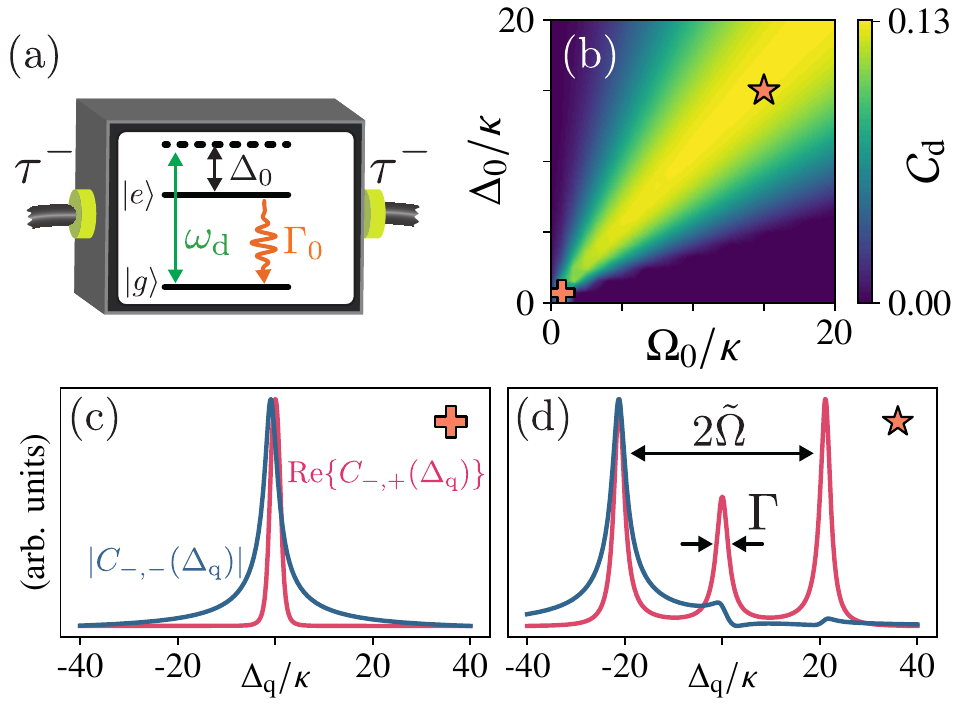}
    \caption{(a) Schematics of a \gls{CPS} consisting of a single \gls{TLS}. The \gls{TLS} is driven by an external field of frequency $\omega_{\rm d}$ and decays into the two waveguides with rates $\kappa$ and into an independent environment with rate $\Gamma_0$. (b) Plot of the distributable entanglement as a function of the drive strength, $\Omega_0$, and the detuning of the drive, $\Delta_{0}$, and for target qubits tuned in resonance with the Mollow sidebands ($\Delta_{\mathrm{q},1}=-\Delta_{\mathrm{q},2}=\tilde{\Omega}$). The two symbols indicate the weak-driving (cross) and strong-driving (star) regimes. For both regimes, the dependence of the two relevant correlation functions $\Re\{C^{11}_{-+}(\Delta_{\rm q})\}\sim N_1$ and $|C^{12}_{--}(\Delta_{\rm q})|\sim |M|$ on the detuning $\Delta_{\rm q}$ is sketched in the respective plots in (c) and (d).
    For better visibility, the correlator spectra have been scaled to the same maximal value.}
    \label{fig:Mollow_0}
\end{figure}

We consider the setup depicted in~\cref{fig:Mollow_0}(a), where the CPS is represented by a single \gls{TLS} with a ground state $|g\rangle$ and an excited state $|e\rangle$, separated by a transition frequency $\omega_0$. To distinguish this \gls{TLS} from the target qubits in the previous discussion, in the following we denote the Pauli matrices acting on the source by $\tau^\alpha$, i.e., $\tau^z=|e\rangle\langle e|-|g\rangle\langle g|$, $\tau^-=|g\rangle\langle e|$, etc. The \gls{TLS} is driven by an external classical field of frequency $\omega_{\rm d}=\omega_0-\Delta_{0}$ and it decays symmetrically into the two waveguides with rates $\kappa_{1,2}=\kappa$ and jump operators $O_{1,2}=\tau^-$. In addition, we account for a decay into a separate third reservoir with rate $\Gamma_0$. In the rotating frame of the drive, the \gls{CPS} is then described by the Lindbladian
\begin{align}
        \mathcal{L}_{\mathrm{CPS}}&=-i\left[\frac{\Delta_{0}}{2}\tau^z+\frac{\Omega_0}{2}\tau^x,\bullet\right] +\Gamma\mathcal{D}[\tau^-],
        \label{Eq:MollowLindbladian}
\end{align}
where $\Gamma=\Gamma_0+2\kappa$ is the total decay rate. 

The Hamiltonian of the driven \gls{TLS} in the first term of Eq.~\eqref{Eq:MollowLindbladian} can be diagonalized by changing to a dressed-state basis
\begin{align}
\begin{aligned}
|\tilde g\rangle &= \cos(\theta/2)\ket{g}-\sin(\theta/2)\ket{e},\\
|\tilde e\rangle &= \cos(\theta/2)\ket{e}+\sin(\theta/2)\ket{g},
\label{Eq:DressedBasis}
\end{aligned}
\end{align}
where $\theta= {\rm atan}(\Omega_0/\Delta_{0})$. This basis transformation introduces the dressed frequency, 
\begin{equation}
\tilde{\Omega}\equiv\sqrt{\Delta_{0}^2+\Omega_0^2},
\end{equation}
as the relevant scale for the coherent dynamics, and we can distinguish two qualitatively different regimes. For weak driving, $\Gamma\gtrsim \tilde{\Omega}$, the \gls{TLS} scatters photons mostly elastically at the drive frequency [see~\cref{fig:Mollow_0}(c)]. In the opposite, strong-driving regime, $\tilde{\Omega}\gg \Gamma$, the emission spectrum
has the characteristic structure of the Mollow triplet~\cite{mollow_power_1969} with three separated scattering peaks at $\omega_\mathrm{d}$ and $\omega_\mathrm{d}\pm \tilde{\Omega}$ [see~\cref{fig:Mollow_0}(d)]. The effective occupation numbers $N_i$ introduced in~\cref{Eq:EffectiveCoefsCPS} are directly proportional to the emission spectrum and thus, as depicted in~\cref{fig:Mollow_0}(c) and (d), they follow the same single-peak or triple-peak structure in the respective regimes. The squeezing correlations, $C^{12}_{--}$, have an asymmetric shape (for $\Delta_0\neq 0$) with a single dominant peak. Note, however, the squeezing parameter $|M|$ is symmetric upon exchanging the detunings $(\Delta_{\mathrm{q},1},\Delta_{\mathrm{q},2})\to (-\Delta_{\mathrm{q},1},-\Delta_{\mathrm{q},2})$.

\subsection{Entanglement} 

For a single TLS, we can readily evaluate the correlation spectra for a given set of parameters  $\Delta_{0}$, $\Omega_0$, and $\Gamma_0$ and optimize the resulting entanglement $\mathcal{C}_\mathrm{d}$ with respect to the qubit detunings $\Delta_{\mathrm{q},i}$.
For $\Gamma_0=0$, the results are shown in~\cref{fig:Mollow_0}(b) and demonstrate a nonvanishing amount of distributable entanglement over a large range of parameters. The maximal concurrence, however, is limited to $\mathcal{C}_\mathrm{d}\approx 0.13$, which is reached for values of $\Delta_{0}\approx \Omega_0$ and for qubits tuned symmetrically to the opposite Mollow sidebands, $\Delta_{\mathrm{q},1}=-\Delta_{\mathrm{q},2}= \tilde{\Omega}$. This parameter regime is consistent with the optimal one found in Ref.~\cite{lopez_carreno_entanglement_2024}. Interestingly, this optimal value of $\mathcal{C}_\mathrm{d}\approx 0.13$ is almost independent of the ratio $\Omega_0/\kappa$, and can be reached even in the weak-driving regime, where the Mollow sidebands are no longer resolved. The entanglement also remains robust when an additional decay channel with $\Gamma_0\neq 0$ is introduced. For example, for $\Gamma_0=\kappa/2$, the optimal entanglement is only reduced to  $\mathcal{C}_\mathrm{d}\approx 0.1$. This remains true as long as $\Gamma_0\lesssim\kappa$ and most of the photons are still emitted into the two waveguides.  In additional numerical simulations (not shown), we also confirm that the entanglement remains robust in the presence of dephasing, as long as the dephasing rate $\Gamma_{\phi}$ of the \gls{TLS} is smaller than $\kappa$.

\subsection{Discussion} 
\begin{figure}
    \centering
    \includegraphics[width=\linewidth]{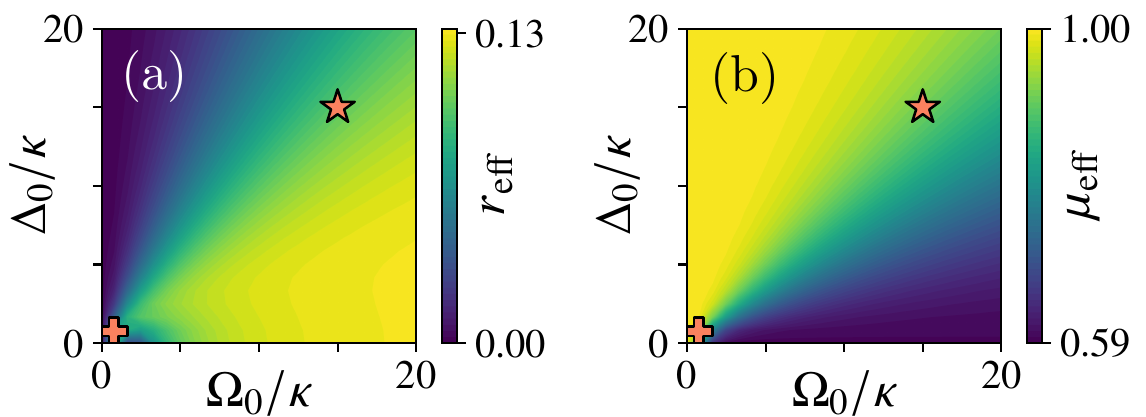}
    \caption{Plots of (a) the effective squeezing strength and (b) the effective purity for a driven \gls{TLS} and for the same parameters as in~\cref{fig:Mollow_0}(b). The opposite trend of those two quantities results in optimal entanglement for $\Delta_{0}\approx\Omega_0$.
    }
    \label{fig:Mollow_1}
\end{figure}

The results for the distributable entanglement shown in~\cref{fig:Mollow_0} and summarized above are largely consistent with the analysis of the entanglement of the emitted photonic modes in Refs.~\cite{lopez_carreno_entanglement_2024,yang_entanglement_2025,vivas-viana_nonclassical_2025}. Our analysis, however, provides a quantifiable measure independent of any additional specifications of the output modes and offers additional physical insights into the observed limitations on entanglement. Specifically, in~\cref{fig:Mollow_1} we show separate plots of the effective squeezing parameter $r_{\rm eff}$ and the effective purity $\mu_{\rm eff}$ under conditions identical to those in~\cref{fig:Mollow_0}(b).  
We see that the squeezing strength increases for a near-resonantly driven \gls{TLS} with $|\Delta_{0}|/\Omega_0\ll1 $, while the purity follows the opposite trend. From these two plots, we can thus understand the maximum of the concurrence at $\Delta_{0}\approx \Omega_0$ as an optimum between the squeezing correlations that are generated in the emitted beams and their purity. In the strong-driving limit $\tilde{\Omega}\to \infty$, the expressions of the coefficients $N_i$ and $M$ [see~\cref{Eq:EffectiveCoefsCPS}] become particularly simple for qubits tuned to the Mollow sidebands. In this case, we find the analytical results
\begin{align}
\begin{aligned}
    N=\frac{4 \sin^4 (\theta )}{4 \cos (2 \theta )+29-\cos (4 \theta )}, \quad M=\frac{\sin^2(\theta )}{\cos (2 \theta )-5}.
\end{aligned}
\end{align}
From these expressions it follows that a maximal squeezing strength of $r_{\mathrm{eff}}=\frac{1}{2}\tanh^{-1}(\frac{1}{4})\simeq 0.12$ is achieved for resonant driving, $\theta=\pi/2$, at which, however, the purity attains its minimal value of $\mu_{\mathrm{eff}}=3/5$.

From this analysis of the effective reservoir parameters, we can also identify the main reasons that limit the distributable entanglement of a \gls{TLS} source to a rather low value of $\mathcal{C}_\mathrm{d}\simeq 0.13$. First of all, the low value of $r_{\rm eff}$, even on resonance, implies that the \gls{TLS} does not emit enough correlated photons. This is in contrast to the case of a parametric amplifier, where there is no fundamental limit on the maximal photon flux. On top of it, in parameter regimes where the squeezing is sufficiently strong, the purity rapidly decreases. 
On the one hand, this degradation arises from the reduced purity of the steady state of the \gls{TLS} itself. On the other hand, the photons from both Mollow sidebands are emitted equally into both waveguides. Therefore, half of the correlations between photons that are not resonant with the target qubits are lost, reducing the effective purity of the remaining photons' state.

\subsection{Finite-bandwidth effects}\label{Sec:FiniteBandwidth}
In our analysis, we focus exclusively on the limit $\gamma/\kappa\rightarrow 0$, in which the bandwidth of the photon source exceeds by far the bandwidth of the target qubits. This limit permits a well-defined quantitative analysis of the CPS, and in previous studies of TMS photon sources~\cite{agusti_long-distance_2022,agusti_autonomous_2023} it has also been shown that this condition maximizes the steady-state entanglement. From a comparison with exact numerical simulations of the full network at finite $\gamma$, we observe the same trend across all investigated configurations and parameter regimes discussed below. For the bare \gls{TLS} source, however, there is a notable and interesting exception to this trend, which occurs for moderate driving strengths, $\kappa \gtrsim \tilde{\Omega}$. In this case, we find that the entanglement between the target qubits---while still being limited---can exceed the bound identified above when $\gamma\approx \kappa$. As we discuss in more detail in~\cref{Sec:NonMarkovian}, under this condition, the whole system relaxes into a more complex tripartite entangled state, in which also the \gls{TLS} of the source becomes entangled with the target qubits.

\section{Cavity-assisted Mollow correlations}  \label{Sec:MollowCavities}
In the previous section, we argued that, for a bare TLS, quantum correlations are degraded because photons from both Mollow sidebands are emitted equally into both waveguides. This suggests that the distributable amount of entanglement can be enhanced by embedding the \gls{TLS} in a structured environment that directs the emission from the Mollow sidebands into opposite directions. This situation can be realized, for example, by channeling the \gls{TLS} emission through narrowband filter cavities with appropriately chosen resonances.

\subsection{Model}
\begin{figure}[t]
    \centering
\includegraphics[width=1\linewidth]{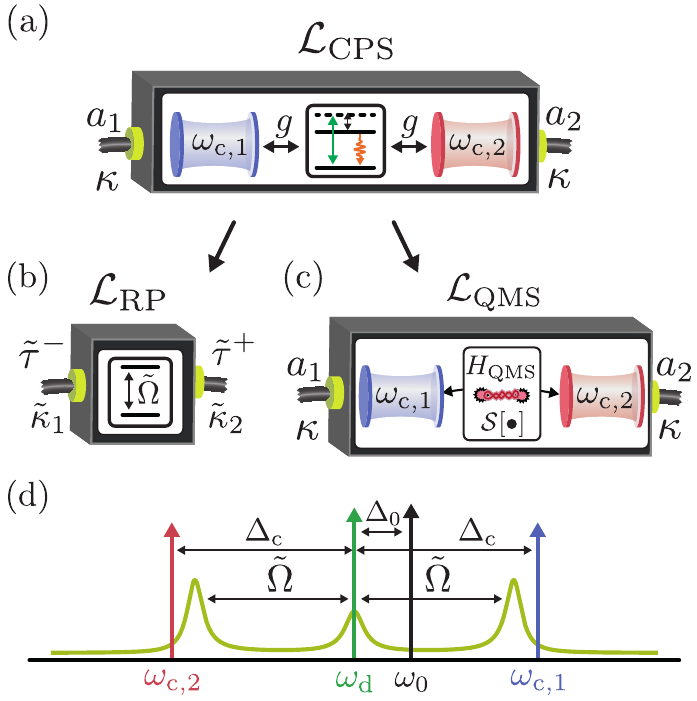}
    \caption{(a) The \gls{CPS} consists of a driven, dissipative \gls{TLS} coherently coupled to two bosonic modes $a_1$ and $a_2$ with strength $g$. We assume that the modes selectively emit into one or the other waveguide, that is $O_1=a_1$, $O_2=a_2$. The \gls{TLS} has the same parameters as shown in \cref{fig:Mollow_0}(a). (b) In the resolved-Purcell regime, $\tilde \Omega\gg \kappa \gg g$, two cavities can be adiabatically eliminated to obtain an effective model for the \gls{TLS}, with jump operators $O_1=\tilde{\tau}^-$, $O_2=\tilde{\tau}^+$ and decay rates $\tilde{\kappa}_1$ and $\tilde{\kappa}_2$. (c) In the opposite regime, where $\Gamma_0 \gg g,\kappa$, one can adiabatically eliminate the \gls{TLS}, and model the remaining dynamics of the two cavities by a quadratic \gls{TMS} Lindbladian. (d) Sketch of the relevant frequencies and detunings in the extended \gls{CPS}: The driven \gls{TLS} emits photons at the two Mollow sidebands at $\omega_\mathrm{d}\pm\tilde{\Omega}$. The symmetrically detuned cavities are located at $\omega_\mathrm{d}\pm \Delta_\mathrm{c}$.}
\label{fig:CPS_Q_cavities}
\end{figure}
To model this scenario, we consider an extended CPS as depicted in~\cref{fig:CPS_Q_cavities}(a). Here, the driven \gls{TLS} is coupled via a Jaynes-Cummings-type interaction to two photonic cavity modes with frequencies $\omega_{\mathrm{c},1}$ and $\omega_{\mathrm{c},2}$ and bosonic annihilation (creation) operators $a_1$ ($a_1^\dag)$ and $a_2$ ($a_2^\dag$). Each cavity decays only into one of the waveguides with rates $\kappa_{1,2}=\kappa$, which we assume to be equal for simplicity. The CPS source is thus described by the two jump operators $O_{1}=a_1$ and $O_2=a_2$ and the Liouvillian
\begin{align}
\begin{aligned}
    \mathcal{L}_{\mathrm{CPS}}=&-i\left[H_{\mathrm{CPS}},\bullet\right]+\Gamma_0\mathcal{D}[\tau^-]+\kappa \mathcal{D}[a_1]+\kappa \mathcal{D}[a_2].
    \label{Eq:MollowCavityBB}
\end{aligned}
\end{align}
In the frame rotating with the driving frequency, the Hamiltonian of the source is given by  
\begin{align}
\begin{aligned}
    H_{\mathrm{CPS}}=\, &\frac{\Delta_{0}}{2}\tau^z+\frac{\Omega_0}{2}\tau^x \\
    &+\sum_{i=1}^2 \left[ \Delta_{\mathrm{c},i}a_i^\dag a_i  +  g\left(a_i\tau^+ +a_i^\dag \tau^-\right)\right],
    \label{Eq:MollowCavityHCPS}
\end{aligned}
\end{align}
where $\Delta_{\mathrm{c},i}=\omega_{\mathrm{c},i}-\omega_{\mathrm{d}}$ are the detunings of the cavity resonances from the drive. For the remainder of this section we assume that $\Delta_{\mathrm{c},1}=-\Delta_{\mathrm{c},2}\equiv \Delta_{\mathrm{c}}\geq 0$ and $\Delta_{\mathrm{q},1}=-\Delta_{\mathrm{q},2}\equiv\Delta_{\mathrm{q}}$ [see~\cref{fig:CPS_Q_cavities}(d)]. We furthermore impose $\Delta_\mathrm{c}=\Delta_\mathrm{q}$ and treat $\Delta_\mathrm{c}$
as a tunable parameter. While none of the following results relies on this precise resonance between the cavities and the target qubits, this condition is usually optimal.

\subsubsection{Dressed-state master equation} \label{Sec:DressedStateME}
In the strong-driving regime, $\tilde{\Omega}\gg g, \kappa, \Gamma_0$, the peaks of the Mollow triplet are well-resolved, and we can efficiently separate the photons from different sidebands. In this regime, it is convenient to express the Hamiltonian $H_{\rm CPS}$ in terms of the dressed basis defined in~\cref{Eq:DressedBasis}. In this basis, the \gls{TLS} Hamiltonian in the first line of~\cref{Eq:MollowCavityHCPS} is diagonal, and we can further express the lowering operator as 
\begin{align}
    \tau^-=\cos^2(\theta/2)\tilde{\tau}^- 
    -\sin^2(\theta/2)\tilde{\tau}^+ +  \frac{\sin(\theta)}{2}\tilde{\tau}^z,
\end{align}
where the $\tilde \tau^\alpha$ denote the Pauli operators for the dressed basis states $|\tilde g\rangle$ and $|\tilde e\rangle$. In the sideband-resolved regime and for $|\Delta_\mathrm{c}+\tilde{\Omega}|\gg g,|\Delta_\mathrm{c}-\tilde{\Omega}|$, a \gls{RWA} can be applied to neglect non-resonant terms. This approximation results in the simplified 
Hamiltonian
\begin{align}
\begin{aligned}
    \tilde{H}_{\mathrm{CPS}} \overset{\text{RWA}}{\approx} &\frac{\tilde{\Omega}}{2}\tilde{\tau}^z+\Delta_\mathrm{c} (a_1^\dag a_1-a_2^\dag a_2)  \\
    &+ \left(g_ca_1^\dag-g_sa_2\right)\tilde{\tau}^- + \text{H.c.},
    \label{Eq:DressedBasisRWA}
\end{aligned}
\end{align}
where $g_c\equiv g\cos^2(\theta/2)$ and $g_s\equiv g\sin^2(\theta/2)$. Under the same conditions and assuming further that $\tilde{\Omega}\gg \Gamma_0$, we can also apply a \gls{RWA} to the dissipator of the \gls{TLS} and approximate the CPS by the dressed-state \gls{ME} with Liouvillian 
\begin{align}
\begin{aligned}
    \tilde{\mathcal{L}}_{\mathrm{CPS}}=&-i\left[\tilde H_{\mathrm{CPS}},\bullet\right]
    +\kappa \mathcal{D}[a_1]+\kappa \mathcal{D}[a_2]\\
    &+\tilde{\Gamma}_{z}\mathcal{D}[\tilde{\tau}^z]+\tilde{\Gamma}_{+}\mathcal{D}[\tilde{\tau}^+]+\tilde{\Gamma}_{-}\mathcal{D}[\tilde{\tau}^-]
     \label{Eq:RWAME}
\end{aligned}
\end{align}
and rates
\begin{align}\label{eq:dressed_dissipation_rates}
\begin{aligned}
    \tilde{\Gamma}_{z}&=\Gamma_0\sin^2(\theta)/4, \\
    \tilde{\Gamma}_{+}&=\Gamma_0\sin^4(\theta/2), \\
    \tilde{\Gamma}_{-}&=\Gamma_0\cos^4(\theta/2).
\end{aligned}
\end{align}
We see that, in this dressed-state representation and for $\tilde \Omega\approx \Delta_\mathrm{c}$, the \gls{TLS} induces transitions that both excite and destroy cavity photons, but with a fixed phase relation. At the same time, the dissipator describing decay in the original basis induces incoherent transitions between dressed states and dephasing. We emphasize that this effective model relies on the \gls{RWA} and does not fully capture the physics of the weak-driving regime. Under strong-driving conditions, however, it greatly simplifies numerical simulations, and throughout the following discussion, we will use the approximations in~\cref{Eq:DressedBasisRWA} and~\cref{Eq:RWAME} for analytic predictions and physical interpretations of the results.

\subsubsection{Parameter regimes}
The inclusion of filter cavities introduces the coupling strength $g$ as an additional coherent frequency scale, which must be compared---together with $\tilde \Omega$---with the relevant dissipation rates $\kappa$ and $\Gamma_0$. To investigate the entangling capability of the CPS in this large parameter space, we focus separately on four limiting cases: (i) In the unresolved Purcell regime, $\kappa\gg \tilde\Omega, g$, the cavity resonances are sufficiently broad to cover the full Mollow spectrum, and the cavities simply enhance the emission into the waveguides. In this limit,  we recover the physics of a bare \gls{TLS} discussed in~\cref{Sec:DrivenTLS} above with a redefined decay rate, $\kappa\rightarrow 4g^2/\kappa$. (ii) In the \gls{RP} regime, $\tilde\Omega\gg \kappa\gg g$, the cavities still primarily channel the decay of the \gls{TLS} into the waveguides, but in a sideband-resolved manner. (iii) In the \gls{QMS} regime, $\Gamma_0 \gg \kappa, g$, the relaxation dynamics of the \gls{TLS} is faster than that of the cavity modes and the \gls{TLS} then acts as an effective TMS reservoir for the slowly decaying cavities. (iv) Finally, in the strong coupling regime, $g\geq \kappa, \Gamma_0$, the coherent Jaynes-Cummings interaction dominates over both dissipation rates and strongly hybridized excitations between the dressed \gls{TLS} states and the cavity modes can emerge.

\begin{table*}
\begin{center}
\begin{tabular}{|c|c|c|c|c|}
\cline{2-5}
\multicolumn{1}{c|}{} 
& \multicolumn{4}{c|}{Parameter regime} \\
\hline
Model
& \begin{tabular}{@{}c@{}}
(i) Unresolved Purcell\\
$\kappa \gg \tilde \Omega,g,\Gamma_0$ 
\end{tabular}
& \begin{tabular}{@{}c@{}}
(ii) Resolved Purcell \\
$\tilde \Omega \gg\kappa \gg g,\Gamma_0$
\end{tabular}
& \begin{tabular}{@{}c@{}}
(iii) \gls{QMS} \\
$\Gamma_0 \gg g,\kappa$
\end{tabular}
& \begin{tabular}{@{}c@{}}
(iv) Strong Coupling \\
 $g \geq \kappa,\Gamma_0$
\end{tabular} \\
\hline
\cref{Sec:DrivenTLS}              & \checkmark &  $\times$ &  $\times$ & $\times$ \\
\arrayrulecolor{gray!80}\hline\arrayrulecolor{black}
$\mathcal{L}_{\rm RP}$    & $\times$   & \checkmark &  $\times$ & $\times$ \\
\arrayrulecolor{gray!80}\hline\arrayrulecolor{black}
$\mathcal{L}_{\rm QMS}$           & $\times$   & $\times$ & \checkmark &(\checkmark) \\
\arrayrulecolor{gray!80}\hline\arrayrulecolor{black}
$\mathcal{L}_{\rm CPS}$           & \checkmark & \checkmark & \checkmark & \checkmark \\
\hline
\end{tabular}
\end{center}
\caption{Summary of the different parameter regimes and applicability of the effective models considered in~\cref{Sec:MollowCavities}. The \gls{QMS} master equation is also valid in the strong-coupling regime, $g\geq \kappa, \Gamma_0$, under additional conditions discussed in~\cref{Sec:StrongCoupling}.}
\label{Tab:EffectiveParamRegimes}
\end{table*}

In~\cref{Tab:EffectiveParamRegimes} we summarize the different parameter regimes and the validity of the exact and approximate models, which we will use in the following discussion to obtain analytical estimates and simplified numerical simulations. Whenever possible, these approximate results are benchmarked against numerical simulations of the full system; however, this becomes rather challenging at large effective squeezing, as the emission spectrum is very sensitive to truncation errors. In the strong-driving regime, $\tilde \Omega\gg g,\kappa$, the \gls{RWA} for the dressed states in~\cref{Eq:DressedBasisRWA} and~\cref{Eq:RWAME} can be used to partially mitigate the numerical costs while still retaining sufficiently high  accuracies.

\subsection{Purcell regime} \label{Sec:MollowCavitiesPurcell}
We first address the Purcell~\cite{purcell_resonance_1946} regime, $\kappa\gg g$, where the cavities primarily enhance the incoherent decay into the waveguide. There are two qualitatively different situations: (i) For $\tilde{\Omega}\ll \kappa$, the cavity modes can be adiabatically eliminated, and we recover the physics of a bare \gls{TLS} coupled to two broadband waveguides with effective decay rates $\tilde{\kappa}= 4g^2/\kappa$. (ii) If $\tilde{\Omega}\gg \kappa$, the adiabatic elimination  of the cavities must be carried out in the dressed basis, which leads to selective enhancement of the emission at the Mollow sidebands. As the first case was already discussed in \cref{Sec:DrivenTLS}, we focus here on the \gls{RP} regime. For simplicity, we assume for the remainder of this discussion that the qubit relaxes only through the cavities, i.e., $\Gamma_0=0$.

\subsubsection{Resolved Purcell master equation}
To model the CPS in the sideband-resolved Purcell regime, it is convenient to work in the dressed-state basis defined in~\cref{Eq:DressedBasis}. After eliminating the fast dynamics of the cavity modes, we obtain a master equation for the reduced state of the driven TLS 
\begin{equation}
\mathcal{L}_{\rm RP} \simeq  -i\left[\frac{\tilde{\Omega}}{2}\tilde{\tau}^z,\bullet\right]+ \tilde{\kappa}_1 \mathcal{D}[\tilde{\tau}_-] + \tilde{\kappa}_2 \mathcal{D}[\tilde{\tau}_+].
\label{Eq:PurcellCPS}
\end{equation} 
Here, the effective Purcell rates are given by~\cite{gonzalez-ballestero_tutorial_2024}
\begin{align}
\begin{aligned}
    \tilde{\kappa}_1=\frac{g_c^2\kappa}{(\Delta_{\mathrm{c},1}-\tilde{\Omega})^2+(\kappa/2)^2},\\
    \tilde{\kappa}_2=\frac{g_s^2\kappa}{(\Delta_{\mathrm{c},2}+\tilde{\Omega})^2+(\kappa/2)^2}.
    \label{Eq:PurcellME}
\end{aligned}
\end{align}
The \gls{CPS} thus corresponds to a driven \gls{TLS} emitting with the dressed qubit operators $\tilde{\tau}^-$ ($\tilde{\tau}^+$) into the first (second) waveguide [see~\cref{fig:CPS_Q_cavities}(b)].
\begin{figure}[t]
    \centering
    \includegraphics[width=\linewidth]{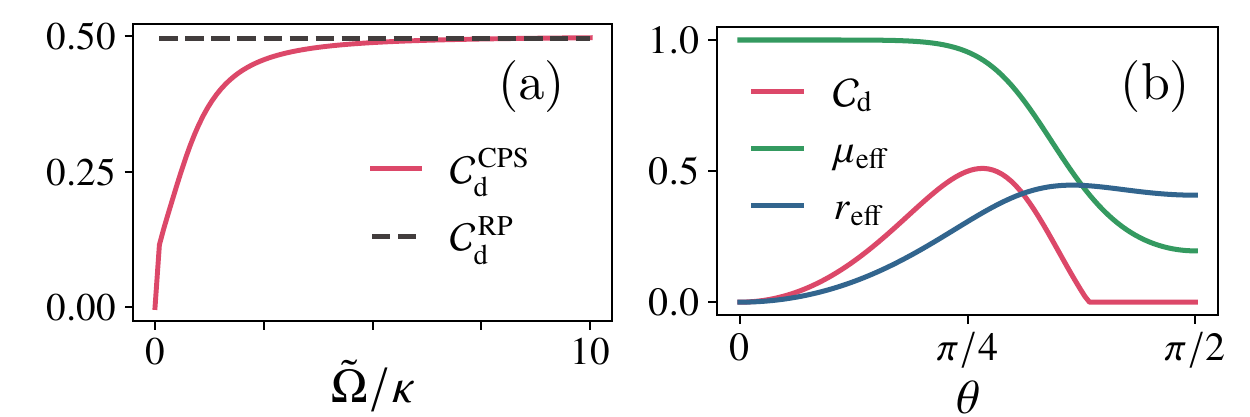}
    \caption{(a) Concurrence $\mathcal{C}_\mathrm{d}$ as a function of the sideband frequency $\tilde{\Omega}$ for fixed $\theta=\pi/4$. (b) Plot of the concurrence and the effective squeezing parameters for varying dressing angle $\theta$ at fixed $\tilde{\Omega}=30\kappa$. In both panels, the solid lines show the results obtained from an exact numerical simulation of~\cref{Eq:MollowCavityBB} with three Fock states per mode and in the bad cavity regime with $\kappa=10g$ and $\Delta_\mathrm{c}=\tilde{\Omega}$.  In (a), the dashed line shows the corresponding results obtained from the approximate RP \gls{ME} in~\cref{Eq:PurcellME}.}
\label{fig:Mollow_and_Cavities_0}
\end{figure}

\subsubsection{Purcell-enhanced entanglement}

In~\cref{fig:Mollow_and_Cavities_0} we show the resulting amount of distributable entanglement in the \gls{RP} regime for $\kappa/g=10$ and varying $\tilde{\Omega}$ and mixing angle $\theta$. In~\cref{fig:Mollow_and_Cavities_0}(a), we first sweep the sideband separation $\tilde{\Omega}$ at a fixed dressing angle $\theta=\pi/4$. The effective \gls{RP} model (black dashed line) matches the predicted concurrence for strong driving fields, $\tilde{\Omega}\gg g$, whereas discrepancies in the weak-driving regime arise from the failure of the \gls{RWA} in~\cref{Eq:DressedBasisRWA}. Overall, we observe that filtering works best when the cavities are well separated, allowing us to select only one of the two sideband peaks.
In panel (b), we sweep the dressing angle $\theta$ for fixed $\tilde{\Omega}=30\kappa$, such that $\Omega_0=\tilde{\Omega}\sin(\theta)$ and $\Delta_{0}=\tilde{\Omega}\cos(\theta)$. As already observed for a single \gls{TLS} in the strong-driving regime, the concurrence vanishes also in the current setting for a resonant drive, $\Delta_{0}\to 0$ ($\mathrm{\theta\to \pi/2}$), 
and the maximal concurrence is reached around $\theta=\pi/4$ ($\Delta_{0}\approx \Omega_0$). 

\subsubsection{Discussion}
Compared to the symmetric decay into both waveguides considered in~\cref{Sec:DrivenTLS}, the crucial difference in the resolved Purcell regime is that the photons from the two sidebands now only illuminate the target qubits with matched frequencies. Therefore, most of the photonic correlations generated by the \gls{TLS} also contribute to the entanglement-stabilization process. 
As can be seen in~\cref{fig:Mollow_and_Cavities_0}(b), when compared to the bare TLS, filtering photons from the sideband results in a slightly higher effective purity $\mu_\mathrm{eff}$ near $\theta\approx \pi/4$. It also increases the effective squeezing parameter $r_\mathrm{eff}$ compared to the unfiltered waveguide scenario. However, even with those improvements, we find that the maximally achievable concurrence between the target qubits is bounded by $\mathcal{C}_\mathrm{d}\approx 0.51$. 
This value can be understood by considering the limit $\tilde{\Omega}\to \infty$, where for cavities tuned to the two Mollow sidebands, the analytical expressions for the effective reservoir parameters in~\cref{Eq:EffectiveCoefsCPS} read
\begin{align}
\begin{aligned}
    N&=\frac{4\sin^4(\theta)}{[3+\cos(2\theta)]^2}, \qquad M=\sqrt{N}.
\end{aligned}
\end{align}
Together with~\cref{Eq:TMSEff} and~\cref{Eq:DefConc}, these expressions indeed predict an optimal value of $\mathcal{C}_\mathrm{d}\approx 0.5$ for the concurrence at around $\theta\approx \pi/4$. In the limit of very strong driving, $\theta\to \pi/2$, the squeezing strength is $r_{\rm eff}\approx 0.4$, but $\mu_\mathrm{eff}=1/5$ is significantly lower than in the case of a \gls{TLS} that emits photons directly into the waveguides without the Purcell filters. This shows that the efficiency of photonic correlations generated by a single \gls{TLS} remains limited by the intrinsic photon-emission process.

\subsection{Qubit-mediated squeezing regime}  \label{Sec:MollowCavitiesFilter}
To overcome the limitations identified in the previous section, we now consider the opposite regime, $\Gamma_0\gg g,\kappa$, in which the \gls{TLS} relaxation dynamics are fast compared to the decay of the filter cavities. In this regime, a simplified description of the CPS can be obtained by adiabatically eliminating the \gls{TLS}, yielding an effective \gls{ME} for the cavity modes~\cite{kustura_effective_2021}.

\subsubsection{Effective cavity master equation}
As a result of this adiabatic elimination, we obtain 
a quadratic Liouvillian for the cavity modes,
\begin{align}
 \mathcal{L}_{\rm QMS}=-i[H_{\mathrm{QMS}},\bullet]+\mathcal{S}[\bullet] + \kappa \sum_i \mathcal{D}[a_i].
    \label{Eq:QMSME}
\end{align}
Here, we have introduced the effective cavity Hamiltonian
\begin{align}
\begin{aligned}
    H_{\rm QMS}=&\Delta_\mathrm{c} (a_1^\dag a_1-a_2^\dag a_2)  \\
    &+\sum_{i,j}\delta_{i,j}a_i^\dag a_j 
    +\left(g_{i,j}a_i a_j+g^*_{i,j}a^\dag_i a^\dag_j\right),
\end{aligned}
\end{align}
where $\delta_{i,j}^*=\delta_{j,i}$ and $g_{i,j}^*=g_{j,i}$, and the dissipative part of the Lindbladian reads
\begin{align}
    \mathcal{S}=\sum_{i,j}&\Big(\Gamma_{i,j}\mathcal{D}[a_i,a_j]+\Gamma^*_{i,j}\mathcal{D}[a_j^\dag,a_i^\dag] \nonumber \\
    &+\gamma_+^{i,j}\mathcal{D}[a_i^\dag,a_j]+\gamma_-^{i,j}\mathcal{D}[a_i,a_j^\dag]\Big).
    \label{Eq:DissipatorMollowRegime}
\end{align}
A detailed derivation of~\cref{Eq:QMSME} and all its parameters is given in~\cref{Sec:AETLS}. Note that the additional decay processes described by $\mathcal{S}$ arise from the interaction with the strongly-dissipative \gls{TLS} that is coupled to its own reservoir. They do not affect the emission of the cavities into the waveguides, which is still described by the bare dissipators $\mathcal{D}[a_1]$ and $\mathcal{D}[a_2]$ in Eq.~\eqref{Eq:QMSME}.

Since the effective ME in~\cref{Eq:QMSME} is quadratic in $a_i$ and $a_i^\dag$, its steady state is Gaussian and can be fully characterized by the first and second moments of the cavity operators. This property allows us to compute the effective coefficients in~\cref{Eq:BBCoefs} without introducing truncation errors (see~\cref{Sec:QRT}). 

\subsubsection{Numerical results}
\begin{figure}[t]
    \centering
    \includegraphics[width=\linewidth]{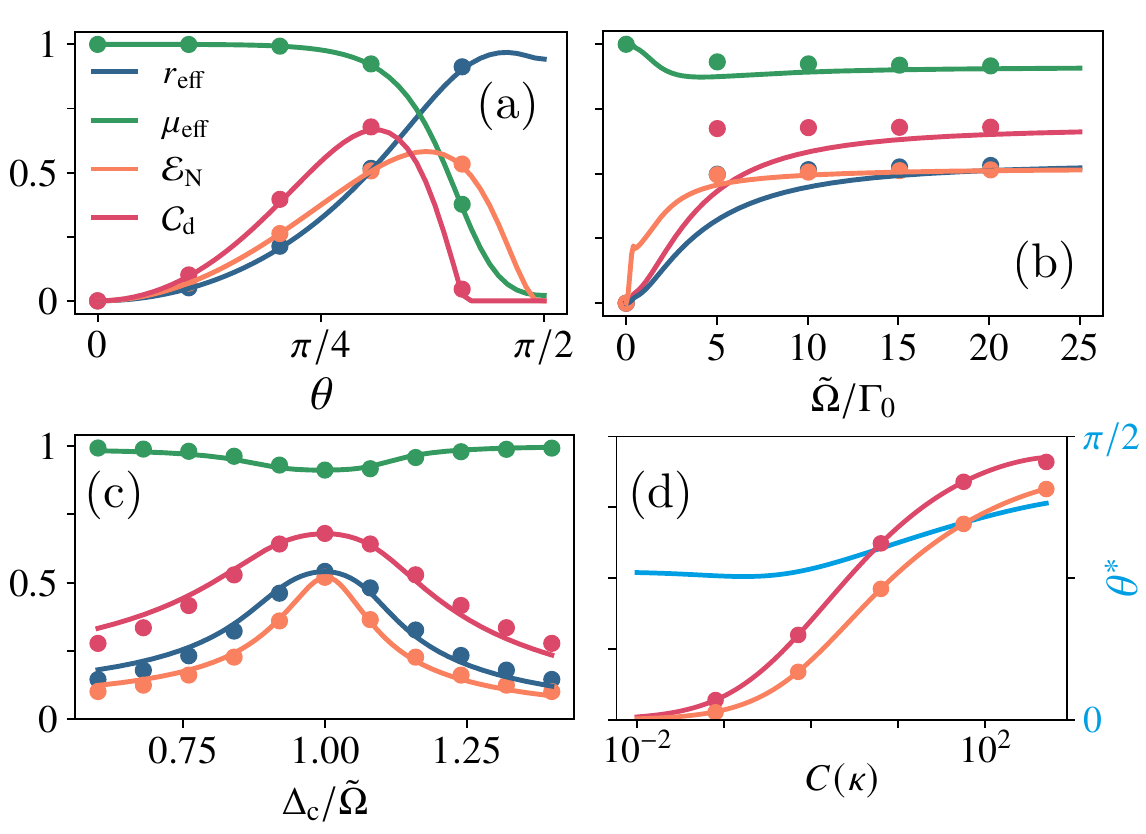}
    \caption{Entanglement distribution in the QMS regime. (a) Concurrence, logarithmic negativity and effective \gls{TMS} parameters as a function of the mixing angle $\theta=\atan(\Omega_0/\Delta_0)$ and for fixed parameters $\tilde{\Omega}=30\Gamma_0$, $\kappa=\Gamma_0/1000$, and $g=\Gamma_0/10$.  (b) Plot of the same quantities as in (a) for a varying sideband parameter $\tilde{\Omega}/\Gamma_0$ and for an optimal dressing angle $\theta^*$ that maximizes the concurrence. In both (a) and (b), the cavities are tuned in resonance with the Mollow sidebands, $\Delta_\mathrm{c}=\tilde{\Omega}$. (c) Plot of the same quantities, but with the cavities and qubits symmetrically detuned from the sidebands, $\Delta_{\rm c}=\Delta_{\rm q}$, and for $\tilde{\Omega}=100\Gamma_0$. (d) Plot of $\mathcal{C}_{\rm d}$ and the entanglement negativity $\mathcal{E}_{\rm N}$ of the cavity modes obtained for an optimal mixing angle $\theta^*$. These quantities are shown as a function of the cooperativity $C=g^2/(\kappa\Gamma_0)$ for fixed $g/\Gamma_0= 0.1$, sideband parameter $\tilde{\Omega}=500\Gamma_0$ and $\Delta_{\rm c}=\tilde \Omega$. 
    In all panels, the solid lines correspond to the results obtained from the effective cavity \gls{ME} in~\cref{Eq:QMSME}, while the dots mark the results  obtained from the dressed-state  \gls{ME} in~\cref{Eq:RWAME} with a truncation of 20 Fock states per mode.}
\label{fig:Mollow_and_Cavities_1}
\end{figure}

In~\cref{fig:Mollow_and_Cavities_1}, we present the results of a numerical exploration of the CPS in the \gls{QMS} regime for a fixed value of $\Gamma_0=10g$
and assuming that both the cavities and the target qubits are tuned to the Mollow sidebands, $\Delta_\mathrm{c}=\tilde{\Omega}$. 
Under this resonance condition, but tuning the ratio $\Omega_0/\Delta_{0}$, we show in~\cref{fig:Mollow_and_Cavities_1}(a) the effective squeezing and entanglement parameters of the CPS for fixed values $\tilde{\Omega}=30\Gamma_0$ and $\kappa=\Gamma_0/1000$. These values translate into a cooperativity of $C=g^2/(\kappa \Gamma_0)=10$, which measures the strength of TLS-induced dissipation processes with respect to the bare cavity decay.  The plot shows four relevant figures of merit: The logarithmic negativity $\mathcal{E}_\mathrm{N}$~\cite{horodecki_quantum_2009} as a measure of entanglement between the two bosonic modes, the effective squeezing parameters $r_\mathrm{eff}$ and $\mu_\mathrm{eff}$, and the resulting distributable entanglement $\mathcal{C}_\mathrm{d}$. For all quantities, we observe excellent agreement between the results obtained from the effective quadratic \gls{ME} (solid lines) and those from full numerical simulations (dots) of \cref{Eq:MollowCavityBB} under a \gls{RWA} and with 20 photon number states per mode.

As in the previous scenarios studied in~\cref{Sec:DrivenTLS} and \cref{Sec:MollowCavitiesPurcell}, the bosonic and the distributable entanglement are both low for $\theta\approx 0$, where the effective squeezing strength is small, and around $\theta\approx \pi/2$, where the effective purity drops significantly. Similar to the Purcell regime, the optimal driving angle $\theta^*$ that maximized the qubit entanglement is close to $\theta^*\approx \pi/4$, but the corresponding concurrence can reach values above the previously encountered bound of $\mathcal{C}_{\rm d}\approx0.5$. In general, the qubit entanglement and the entanglement between bosonic modes (quantified by the logarithmic negativity $\mathcal{E}_\mathrm{N}$) follow a similar overall trend. They are, however, maximized at different driving parameters, and there is no one-to-one correspondence between them. Indeed, there are regimes in which the cavity modes are in an entangled state while the distributable entanglement between the qubits vanishes [see ~\cref{fig:Mollow_and_Cavities_1}(a) in the region below $\theta \approx \pi/2$]. The emission of entangled photon pairs from the driven \gls{TLS} is thus a necessary but not a sufficient condition for autonomous entanglement distribution applications.

\subsubsection{Sideband resolution and robust entanglement}

In~\cref{fig:Mollow_and_Cavities_1}(b), we vary the separation between the Mollow sidebands, $\tilde{\Omega}$, while keeping the resonance condition $\Delta_\mathrm{c}=\tilde{\Omega}$ and the cooperativity $C=g^2/(\kappa \Gamma)=10$ fixed. 
The plot shows the key figures of merit discussed above for the optimized mixing angle $\theta^*$. We see that both concurrence and negativity initially increase with larger sideband separation, before they eventually saturate. 
This dependence confirms that a sideband resolution of about $\tilde{\Omega}/\Gamma_0\gtrsim 10$ is necessary to obtain significant amounts of entanglement. This plot also illustrates the validity of the \gls{RWA} in our numerical simulations. It naturally fails in the weak-driving regime but accurately captures all the figures of merit in the sideband-resolved regime of interest. 

Lifting the constraint of tuning the cavities exactly in resonance with the Mollow sidebands, in~\cref{fig:Mollow_and_Cavities_1}(c), we sweep the cavity detuning $\Delta_\mathrm{c}$ across the ideal value at $\tilde{\Omega}=100 \Gamma_0$ and optimize the dressing angle $\theta^*$ for each detuning separately. While both the concurrence and the entanglement negativity are reduced away from resonance, we find that the distributable entanglement is considerably less affected by a detuning when compared to the entanglement between the cavity modes. Further, the width of the entanglement peak is determined by $\tilde \Omega$, and not by the cavity linewidth $\kappa$. Both features make entanglement distribution rather robust and insensitive to experimental imperfections.

\subsection{Optimal entanglement: from weak to strong coupling} \label{Sec:MollowCavitiesStrongCoupling}
In the previous numerical examples, we have already seen that in the QMS regime, the distributable amount of entanglement can exceed the bound of $0.5$, which we identified for the concurrence in the Purcell regime. In~\cref{fig:Mollow_and_Cavities_1}(d) we address this point more systematically and plot the maximally achievable concurrence as a function of the cooperativity, assuming a fixed coupling $g=\Gamma_0/10$ and well-resolved, resonant interactions, $\Delta_\mathrm{c}=\tilde{\Omega}=500\Gamma_0$. We see that in contrast to the Purcell limit, the distributable entanglement can reach values arbitrarily close to unity, as long as the cavity decay rate is sufficiently low.  Intuitively, we can understand this enhancement by noting that for $C\gg1$, the \gls{TLS} can emit many correlated photon pairs, which accumulate in the cavities before they decay and result in much stronger effective squeezing correlations. In parameter regimes accessible to exact numerical calculations, we obtain values of the concurrence up to $\mathcal{C}_\mathrm{d}\approx 0.93$. However, according to predictions from the quadratic ME in~\cref{Eq:QMSME}, the limit $\mathcal{C}_\mathrm{d}\rightarrow 1$ can be systematically approached when $\kappa\rightarrow 0$, while remaining deep in the sideband-resolved regime to reduce the residual impact of counter-rotating terms in the steady state. 

\begin{figure}
    \centering
    \includegraphics[width=0.7\linewidth]{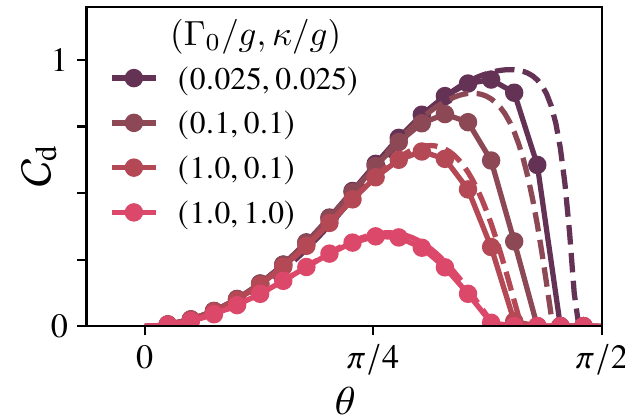}
    \caption{Plot of the distributable entanglement as a function of the mixing angle $\theta$ and for $\tilde{\Omega}=\Delta_\mathrm{c}=75g$. The different colors correspond to different values of $(\Gamma_0/g,\kappa/g)$, which range from the weak-coupling regime, $g\ll \kappa, \Gamma_0$, to the strong-coupling regime, $g\gg \kappa, \Gamma_0$. The dashed lines correspond to the results obtained from the effective \gls{QMS}  \gls{ME}, derived in~\cref{Eq:QMSME} under weak-coupling conditions. The dots and solid lines show the numerical results obtained under a \gls{RWA} approximation in the Bogoliubov basis, as in~\cref {Eq:BogoliubovRWA}, for a truncation of 12 photons per mode.}
    \label{fig:Mollow_and_Cavities_3}
\end{figure}

Apart from lowering $\kappa$, the cooperativity can also be enhanced by increasing $g$. This intuition, however, is based on the QMS ME derived in~\cref{Eq:QMSME} under the weak-coupling condition $g\ll \Gamma_0$. To go beyond this constraint, in~\cref{fig:Mollow_and_Cavities_3} we show exact numerical results for the concurrence $\mathcal{C}_\mathrm{d}$ as a function of $\theta$ and for different values of $\kappa$ and $\Gamma_0$, ranging from the weak- to the strong-coupling regime. While in the latter case, a strong hybridization between the \gls{TLS} and the cavity modes, and thus a saturation or degradation of the squeezing effect, is expected, we observe the opposite. The entanglement improves for couplings well above $g\approx \Gamma_0$, and even under strong-coupling conditions, the numerical results closely follow the predictions from the QMS ME.

In the following section, we discuss this puzzling behavior in more detail. For now, we conclude that the distributable entanglement is optimized in the regime $C\gg1$, where squeezing correlation can accumulate in the cavities before decaying into the waveguides. Optimal entanglement requires well-resolved Mollow sidebands in resonance with the cavity modes and is achieved most efficiently in the strong-coupling regime $g\gg \Gamma_0,\kappa$.

\section{Correlations, squeezing and entanglement: A dressed-state analysis} \label{Sec:DressedStateAnalysis}  
 The exploration of different parameter regimes in the previous section revealed that the QMS regime is optimal for entanglement distribution, as only in this regime sufficiently large values of the effective squeezing parameter $r_{\rm eff}$ (while maintaining high purity $\mu_{\rm eff}$) can be achieved. This study, however, does not immediately explain the absence of entanglement for a resonantly driven TLS, where photons at the Mollow sidebands are still emitted in correlated pairs~\cite{arnoldus_photon_1984}. Furthermore, it does not clarify the unexpected entanglement optimization under strong-coupling conditions.

To obtain additional insights into the origin of the squeezing correlations in the QMS and strong-coupling regimes, we focus on the resolved-sideband limit, where we can use the effective dressed-state ME discussed in~\cref{Sec:DressedStateME} to simplify our analysis. Specifically, after making the \gls{RWA} and moving into a rotating frame with respect to the bare cavity Hamiltonian, $\Delta_\mathrm{c}(a_1^\dag a_1-a_2^\dag a_2)$, the Hamiltonian in~\cref{Eq:DressedBasisRWA} can be written as
\begin{align}
    \tilde{H}_\mathrm{CPS}=\frac{\tilde \Delta_z}{2}\tilde{\tau}^z
    +g_\mathrm{B}(\tilde{\tau}^-b_1^\dag +\tilde{\tau}^+b_1).
    \label{Eq:BogoliubovRWA}
\end{align}
Here, $\tilde \Delta_z = \tilde \Omega-\Delta_\mathrm{c}$ is the detuning between the Mollow sidebands  and the respective cavity frequencies, and we have introduced the Bogoliubov modes
\begin{align} \label{Eq:BogoliubovBasis}
    b_1 &=S^\dag(r_\theta) a_1 S(r_\theta) = \cosh(r_\theta)a_1-\sinh(r_\theta)a_2^\dag , \\
    b_2 &=S^\dag(r_\theta) a_2 S(r_\theta) = \cosh(r_\theta)a_2-\sinh(r_\theta)a_1^\dag ,   
\end{align}
with the squeezing operator $S(\chi)$ defined in~\cref{Eq:TMSOp}. The squeezing parameter $r_{\theta}$ and the coupling $g_\mathrm{B}=\sqrt{g_c^2-g_s^2}=g \sqrt{\cos(\theta)}$ are both functions of the mixing angle $\theta$. They are related by $\tanh(r_\theta)=\tan^2(\theta/2)$.

\subsection{Weak-coupling regime} \label{Sec:WeakCoupling}
In the Bogoliubov representation, we see that only the mode $b_1$ is coupled to the TLS. Therefore, after adiabatically eliminating the \gls{TLS} under the condition $\Gamma_0\gg g$, we obtain a quadratic Lindbladian of the form 
\begin{align}
\begin{aligned}
\tilde{\mathcal{L}}_{\rm QMS}=&\;-i[\tilde{H}_{\rm QMS},\bullet]+\Gamma_{\rm c} \mathcal{D}[b_1] + \Gamma_{\rm h} \mathcal{D}[b_1^\dag]\\
 &\;+ \kappa \sum_i \mathcal{D}[a_i].    
\end{aligned}
\label{Eq:EffectiveCavityLindbladian}
\end{align}
The first term, $\tilde{H}_{\rm QMS}\sim  b_1^\dag b_1$, accounts for a frequency shift of the active Bogoliubov mode, which, in essence, translates into a TMS interaction,
\begin{align} \label{Eq:CoherentTMSHamiltonian}
    \tilde{H}_{\rm QMS}\approx  g_{\mathrm{TMS}} \left(a_1^\dag a_2^\dag+ a_1a_2\right),
\end{align}
when expressed in terms of the original modes. The corresponding squeezing strength is 
\begin{align}
        g_{\mathrm{TMS}}&=g_cg_s\left(\frac{\tilde \Delta_z }{\tilde{\Delta}_z^2+\Gamma^2_\perp} \right)\langle \tilde{\tau} ^z\rangle_{\rm ss},
        \label{Eq:CoherentTMS}
\end{align}
where $\Gamma_{\perp}= 2\tilde{\Gamma}_z+(\tilde{\Gamma}_++\tilde{\Gamma}_-)/2$ [see~\cref{eq:dressed_dissipation_rates}]. In addition to this Hamiltonian correction, the coupling to the dressed \gls{TLS} induces both cooling and heating terms for the active Bogoliubov mode, with rates 
\begin{align}
    \begin{aligned}
        \Gamma_{\rm c}&=g_\mathrm{B}^2\left(\frac{\Gamma_\perp }{\tilde{\Delta}_z^2+\Gamma^2_\perp} \right) \left(1-\langle \tilde{\tau} ^z\rangle_{\rm ss}\right), \\
        \Gamma_{\rm h}&=g_\mathrm{B}^2\left(\frac{\Gamma_\perp }{\tilde{\Delta}_z^2+\Gamma^2_\perp} \right)\left(1+\langle \tilde{\tau} ^z\rangle_{\rm ss}\right).
    \end{aligned}
    \label{Eq:RWAEffectiveCoefs}
\end{align}
By writing $\Gamma_{\rm h}/\Gamma_{\rm c}= N_{\rm th}/(N_{\rm th}+1)$, we can characterize this bath in terms of an overall damping rate $\Gamma_{\theta}=\Gamma_{\rm c}-\Gamma_{\rm h}$ and an effective thermal occupation number $N_{\rm th}$. From the population imbalance between the dressed states, 
\begin{align}
    \langle \tilde{\tau} ^z\rangle_{\rm ss}=\frac{\tilde{\Gamma}_+-\tilde{\Gamma}_-}{\tilde{\Gamma}_-+\tilde{\Gamma}_+}= -\frac{\cos^4(\theta/2)- \sin^4(\theta/2)}{\sin^4(\theta/2)+\cos^4(\theta/2)},
\end{align}
we obtain, after some manipulations, $N_{\rm th}=\sinh^2(r_{\theta})$ and  $\Gamma_{\theta}=-2g^2 \cos(\theta)\langle \tilde{\tau} ^z\rangle_{\rm ss}/\Gamma_\perp$ on resonance.

\subsubsection{Coherent squeezing}

\begin{figure}[t]
    \centering
    \includegraphics[width=\linewidth]{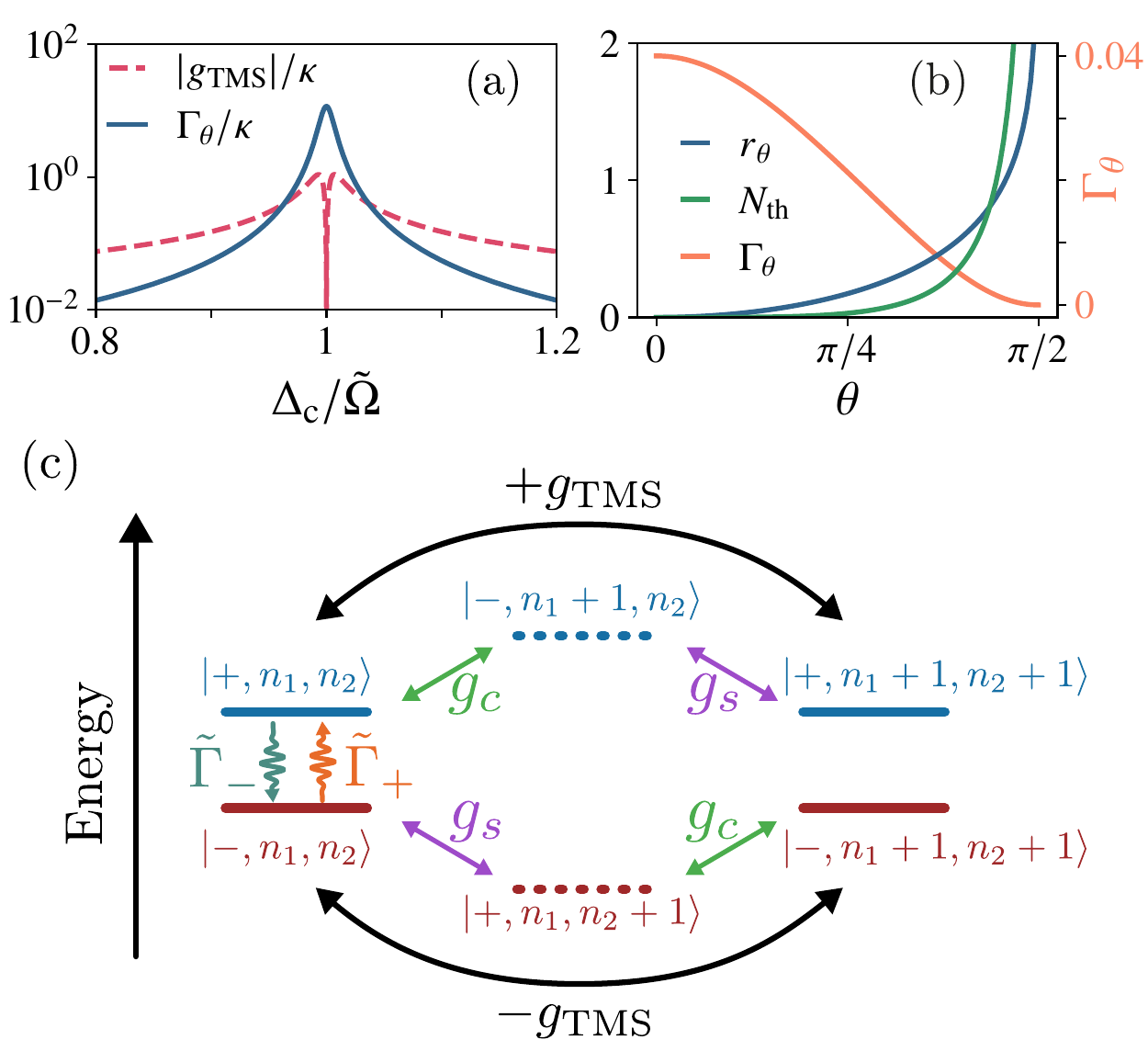}
    \caption{(a) Comparison between the coherent squeezing strength $g_{\rm TMS}$ and the dissipation rate $\Gamma_\theta$, which appear in the weak-coupling \gls{ME} in~\cref{Eq:EffectiveCavityLindbladian}. Both quantities are plotted as a function of the cavity detuning for fixed values of $\theta=\pi/3$, $\tilde{\Omega}=100\Gamma_0$, and $g=\Gamma_0/10$. (b) Plot of the parameters $r_\theta$, $N_{\rm th}$ and $\Gamma_\theta$ that determine the effective evolution of the Bogoliubov modes, see~\cref{Eq:BogoThermal}. For $\theta\to \pi/2$, the squeezing parameter $r_\theta$ is maximized, but the thermal population $N_{\rm th}$ diverges and the effective cooling rate vanishes simultaneously. (c) Level diagram illustrating the two virtual two-photon transitions that contribute to the coherent TMS interaction in~\cref{Eq:CoherentTMSHamiltonian}. The two paths starting from different dressed states contribute with equal magnitude but opposite sign to the \gls{TMS} interaction. }
    \label{fig:Mollow_and_Cavities_2}
\end{figure}

In~\cref{fig:Mollow_and_Cavities_2}(a), we plot the dependence of $g_{\rm TMS}$ and $\Gamma_{\theta}$ as a function of $\Delta_\mathrm{c}$ and for a fixed mixing angle $\theta$. While the coherent squeezing interaction is present also for far detuned cavities, it is---compared to the induced dissipation---less relevant when the cavities are tuned near the Mollow sidebands and vanishes exactly on resonance, where $\tilde \Delta_z=\tilde \Omega-\Delta_\mathrm{c}=0$. The coherent squeezing interaction is also proportional to the population imbalance $\langle \tilde{\tau} ^z\rangle_{\rm ss}$, which vanishes independently of $\tilde \Delta_z$ when the \gls{TLS} is driven on resonance, $\theta=\pi/2$. As depicted in~\cref{fig:Mollow_and_Cavities_2}(c), this cancellation can be intuitively understood as follows. While a two-photon transition via a nonresonant, intermediate dressed state, for example,
\begin{equation}
|+,n_1,n_2\rangle \rightarrow |-,n_1+1,n_2\rangle \rightarrow |+,n_1+1,n_2+1\rangle, 
\end{equation}
induces effective squeezing correlations, the process
\begin{equation}
|-,n_1,n_2\rangle \rightarrow |+,n_1,n_2+1\rangle \rightarrow |-,n_1+1,n_2+1\rangle, 
\end{equation}
results in the same correlated two-photon process, but with an opposite sign (see also Ref.~\cite{sanchez_munoz_simulating_2020} for a related discussion). Therefore, for a resonantly driven TLS, where both dressed states are populated equally, the squeezing interaction vanishes, even though photons at the Mollow sidebands are still emitted in pairs. 

\subsubsection{Dissipative squeezing}
On resonance, the coherent coupling is negligible, and the state of the two filter cavities is mainly determined by the dissipative terms in~\cref{Eq:EffectiveCavityLindbladian}. These include, in particular, the coupling of the Bogoliubov mode $b_1$ to an effective thermal environment with rate $\Gamma_{\theta}$. This process competes with the bare decay of the original modes into the two waveguides. In the limit  $\kappa\rightarrow 0$, the Bogoliubov mode $b_1$ relaxes into a thermal state $\rho_{{\rm th},1}$, but the mode $b_2$ is undetermined. For a nonvanishing but small $\kappa \ll \Gamma_{\theta}$, we can still approximate the combined state of the cavities as a product, $\rho(t)\approx \rho_{{\rm th},1}\otimes \rho_2(t)$, and determine the slow dynamics of the second Bogoliubov modes by 
\begin{equation}
\dot \rho_2 \approx  \kappa \sum_{i=1,2} {\rm Tr}_{b_1}\left\{ \mathcal{D}[a_i]   (\rho_{{\rm th},1}\otimes \rho_2)\right\}.
\end{equation}
Based on this factorized ansatz, we find that the effective evolution of the cavities is approximately given by the Liouvillian 
\begin{align}
\begin{aligned}
    \tilde{\mathcal{L}}_{\rm QMS}\approx\, &\Gamma_{\theta} (N_{\rm th}+1) \mathcal{D}[b_1]+\Gamma_{\theta} N_{\rm th}\mathcal{D}[b_1^\dag]\\
    &+\kappa (N_{\rm th}+1) \mathcal{D}[b_2]+\kappa N_{\rm th}\mathcal{D}[b_2^\dag],
\end{aligned}
\label{Eq:BogoThermal}
\end{align}
i.e., both Bogoliubov modes couple to independent thermal reservoirs with the same equilibrium occupation $N_{\rm th}$, but different rates, $\Gamma_\theta\gg \kappa$.  In the basis of the original cavity modes, 
the combined steady state of the cavities,
\begin{equation}\label{Eq:CavitySteadyState}
\rho_{\rm cav}^{\rm ss}= S(r_\theta) ( \rho_{{\rm th},1}\otimes \rho_{{\rm th},2}) S^\dag(r_\theta),
\end{equation}
is a thermal TMS state with squeezing parameter $r_\theta$ and thermal population $N_\mathrm{th}$, as shown in~\cref{fig:Mollow_and_Cavities_2}(b). 

\subsubsection{Squeezing vs entanglement}
Based on the ME in~\cref{Eq:BogoThermal} for the Bogoliubov modes, we can understand the absence of squeezing correlations for $\theta\rightarrow \pi/2$ from the diverging thermal occupation number $N_{\rm th}$, which in turn traces back to a vanishing population imbalance between the dressed states. Therefore, similar to the coherent interactions discussed above, strong dissipative squeezing processes can be induced by both dressed states, but the effects cancel each other when their steady-state populations are about equal, $ \langle \tilde{\tau} ^z\rangle_{\rm ss}\approx 0$.   

Because of the relation $N_{\rm th}=\sinh^2(r_{\theta})$, the purity of the steady-state in~\cref{Eq:CavitySteadyState} already drops below the entanglement threshold of 1/3 for rather modest squeezing values of $r_\theta \approx 0.6$. To explain the much higher levels of the distributable entanglement observed in simulations, it is important to keep in mind that $\mathcal{C}_{\rm d}\equiv \mathcal{C}_{\rm d}(r_{\rm eff},\mu_{\rm eff})$ depends on effective parameters derived from correlation functions rather than the steady state. For example, according to~\cref{Eq:EffectiveCoefsCPS}, the parameter $M$ is determined by the spectrum of correlations of the form 
\begin{equation}
\langle a_1(\tau)a_2\rangle \sim  \langle b_1(\tau)b_1^\dag\rangle + \langle b_2^\dag(\tau)b_2\rangle.
\label{Eq:CorrelatorBogoliubov}
\end{equation}
Given the very different decay rates, the dominant contribution to the spectrum then arises from $\langle b_2^\dag(\tau)b_2\rangle\sim e^{-\kappa \tau/2}$, which decays on a much longer timescale than the correlations of the first Bogoliubov mode, $\langle b_1(\tau)b_1^\dag\rangle\sim e^{-\Gamma_{\theta} \tau/2}$. 
This picture is confirmed by an exact computation of $N_i$ and $M$ starting from~\cref{Eq:EffectiveCavityLindbladian}, 
from which we obtain 
\begin{align}
\begin{aligned}
N_i&=\frac{\sinh^2(2r_\theta)}{(1+\kappa/\Gamma_{\theta})}, \qquad M=\frac{\sinh(4r_\theta)}{2(1+\kappa/\Gamma_{\theta})},
\end{aligned}
\end{align}
and $r_{\rm eff}$, $\mu_{\rm eff}$ and $\mathcal{C}_{\rm d}$ 
through~\cref{Eq:TMSEff} and~\cref{Eq:DefConc}. In the limit of a large cooperativity, $C\sim \Gamma_\theta/\kappa \gg1$, we find, up to the lowest non-trivial order in $\kappa$, 
\begin{align}
\begin{aligned}
r_{\mathrm{eff}}& \simeq 2r_\theta- \frac{\kappa}{2\Gamma_\theta} \sinh(4r_\theta),\\
\mu_{\rm eff}
&\simeq
1-
4
\frac{\kappa}{\Gamma_\theta}
\sinh^2(2r_\theta).
\end{aligned}
\label{Eq:Analyticalrmu}
\end{align}
We see that a maximal amount of distributable entanglement, $\mathcal{C}_{\rm d}\rightarrow 1$, is reached for sufficiently small $\kappa$, despite the fact that the cavities relax into a highly mixed steady state. The main mechanism that underlies this behavior is that one---and only one---of the Bogoliubov modes is overdamped through a resonant coupling to the dressed states. Importantly, for $\theta\to \pi/2$ we also find that $\Gamma_{\theta}\to 0$, and there is still no entanglement for a \gls{TLS} driven exactly on resonance. For $\theta=\pi/2$ the cavities remain in an exactly factorizable state for arbitrary values of $\kappa$.

\subsection{Strong-coupling regime} \label{Sec:StrongCoupling}
\begin{figure}[t]
    \centering
    \includegraphics[width=\linewidth]{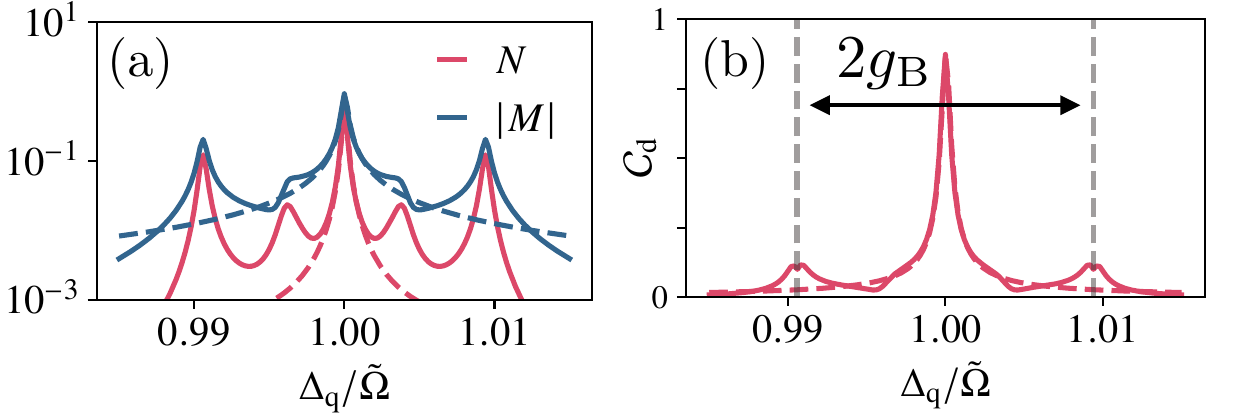}
    \caption{Dependence of (a) the parameters $N$ and $M$ and (b) the concurrence $\mathcal{C}_{\rm d}$ on the symmetric qubit detuning $\Delta_{\rm q}$ in the vicinity of the Mollow sideband. For both plots, strong-coupling conditions with  $\kappa=\Gamma_0=g/40$, $\Delta_{\mathrm{c},1}=-\Delta_{\mathrm{c},2}\equiv \tilde{\Omega}=75g$, and a fixed dressing angle $\theta=\pi/3$ have been assumed. The solid lines are computed from~\cref{Eq:BogoliubovRWA}, the dashed lines correspond to the results obtained from the weak-coupling ME in~\cref{Eq:EffectiveCavityLindbladian}. In both plots, the mode splitting between the $b_1$ mode and the \gls{TLS} leads to the appearance of additional side peaks, but the central peak at $\Delta_{\mathrm{q}}=\tilde{\Omega}$ is still accurately captured by the weak-coupling model.}
    \label{fig:Mollow_and_Cavities_4}
\end{figure}
In the strong coupling regime, but still assuming a sufficiently large separation of the Mollow sidebands, i.e., for $\tilde \Omega\gg g \gg \Gamma_0,\kappa$, the reduced Hamiltonian in~\cref{Eq:BogoliubovRWA} still applies, but neither the \gls{TLS} nor the cavity modes can be adiabatically eliminated. Instead, for $\tilde{\Delta}_z\approx 0$, the Jaynes-Cummings interaction leads to a strong hybridization between the \gls{TLS} and the Bogoliubov mode $b_1$. Thus, we expect the appearance of additional resonances of the combined system at frequencies $\tilde \Omega \pm  g_\mathrm{B}$
around the bare Mollow sidebands. In~\cref{fig:Mollow_and_Cavities_4} we show that these sidebands are indeed visible in the correlation functions that determine $N$ and $M$ and also show up in the concurrence. However, for qubits tuned to the central peak at the Mollow sideband, $\Delta_{\rm q}=\tilde \Omega$, the properties of the spectral functions are essentially unaffected by this Rabi-splitting and are still very accurately modeled by the weak-coupling \gls{ME} derived in~\cref{Eq:EffectiveCavityLindbladian} above (see dashed lines).

The puzzling observation, namely that the strong hybridization between the \gls{TLS} and the cavity modes does not reduce the distributable entanglement, can again be understood from the decomposition of the correlation function in terms of the two Bogoliubov modes, as in~\cref{Eq:CorrelatorBogoliubov}. Here we have argued that only the $b_2$ mode contributes significantly to these correlations, because the $b_1$ mode is heavily damped. Similarly, in the strong-coupling regime, contributions from the $b_1$ mode are shifted to the split sidebands at $\tilde \Omega \pm  g_\mathrm{B}$  and again only the unperturbed $b_2$ mode contributes to the squeezing correlations at the Mollow resonance. This explains why in both coupling regimes, a similar behavior for the distributable entanglement can be observed when either $\kappa\ll \Gamma_\theta$ or $\kappa \ll g$.

\section{Conclusions and Outlook} \label{Sec:Conclusion}
In summary, we have analyzed the use of a single driven \gls{TLS} as a source of quantum-correlated photons for autonomous entanglement distribution applications. For this purpose, we have introduced the distributable entanglement $\mathcal{C}_{\rm d}$ as a quantitative measure to evaluate the suitability of a single \gls{TLS} as a minimal resource for this type of application. From this analysis, we found — consistent with previous related work — that photons emitted at the two opposite Mollow sidebands of a strongly driven \gls{TLS} exhibit useful entanglement. The total amount of entanglement, however, is limited by the maximal amount of emitted photons and by the fact that half of the correlated photons emitted into the wrong Mollow sidebands are lost. 

To overcome these limitations, we have proposed embedding the \gls{TLS} in a structured environment, where the emission from different sidebands is channeled into opposite directions via additional resonant modes. Note that such a scenario differs from simply filtering the scattered output fields~\cite{sanchez_munoz_violation_2014,lopez_carreno_entanglement_2024,vivas-viana_nonclassical_2025,yang_entanglement_2025} and can indeed boost the distributable entanglement up to its maximal value. A detailed analysis of an extended CPS with a driven \gls{TLS} and two cavities revealed many interesting and unexpected properties of this device, which we explained in terms of different approximations, each valid under different coupling and driving conditions. From this analysis, we found that the optimal amount of distributable entanglement is achieved under both strong-driving and strong-coupling conditions.  

The theoretical findings presented in this work are completely general and applicable to any physical platform where a strongly driven \gls{TLS} can be efficiently coupled to cavities and waveguides. For example, in the field of superconducting quantum circuits~\cite{blais_circuit_2021}, both the generation of entangled photon pairs from a driven qubit~\cite{yang_entanglement_2025} and various remote entanglement stabilization schemes~\cite{shah_stabilizing_2024,irfan_autonomous_2025,andres-juanes_entangling_2025}
have already been experimentally demonstrated. These setups can be readily combined and complemented with additional $LC$-resonators to explore all the different configurations and parameter regimes considered in this work.

In terms of applications, however, the current analysis is most relevant for systems where no other well-controlled sources of entangled photons are available. Natural platforms in this context are quantum networks with spin qubits and phononic waveguides~\cite{lemonde_phonon_2018,arrazola_high-fidelity_2024}. It has already been shown that electronic spin qubits associated with, for example, SiV centers in diamond~\cite{hepp_electronic_2014}
can be efficiently coupled to confined acoustic modes~\cite{lee_topical_2017,joe_observation_2025},
which can further be embedded into phononic waveguides and structured networks~\cite{lemonde_phonon_2018,arrazola_high-fidelity_2024,kuzyk_scaling_2018,neuman_phononic_2021,dong_unconventional_2021,shandilya_diamond_2022}. However, compared to other quantum technology platforms, controlled interfaces between spins and phonons or efficient sources of \gls{TMS} states are not yet available in these systems, making the distribution of entanglement via an auxiliary, driven \gls{TLS} an experimentally attractive approach. A crucial benefit of this scheme is that the correlated phonons can be produced by any \gls{TLS} defect that is strain-coupled to phonon modes, without requiring the full control and readout capabilities of the target qubits. This increases the possible options for building such a device, while static strain~\cite{meesala_strain_2018} or Zeeman tuning and strain-induced Rabi frequencies of up to $\sim1$ GHz~\cite{cornell_all-mechanical_2025} provide enough flexibility to match the Mollow sidebands with the targeted qubit frequencies.  Similar benefits may also arise in solid-state systems operated in the optical regime, where many \gls{TLS} are naturally available and can be turned into efficient sources of entanglement by embedding them into cavities or photonic crystal structures~\cite{lodahl_interfacing_2015,gonzalez-tudela_light_2024}. 

The code and data used to produce the results in this paper are openly available in~\cite{gigon_2026}.

\acknowledgments
We thank L. Garbe, X. H. H.  Zhang, M. Fadel, L. Schamriß, and A. Vivas-Via\~na for stimulating comments and discussions. This was supported by the Swiss National Science Foundation through Project Nr. CRSII 222812/1. A. P.-R. acknowledges support from the European Union’s Marie Skłodowska-Curie Actions under grant agreement No. 101204967 (FTMcQED). J. A. acknowledges support from the QUANTERA project MOLAR with reference PCI2024-153449, funded by MICIU/AEI/10.13039/501100011033 and the European Union. This research is part of the Munich Quantum Valley, which is supported by the Bavarian state government with funds from the Hightech Agenda Bayern Plus.

\appendix

\section{Adiabatic elimination of CPS} 
\label{Sec:AdiabaticEliminationCPS}
Here we explicitly carry out the adiabatic elimination of the \gls{CPS} under the assumptions mentioned in the main text: (i) The total Lindbladian is of the form $\mathcal{L}=\mathcal{L}_{\mathrm{CPS}}+\mathcal{L}_{\rm Q}+\mathcal{L}_{\rm I}$, (ii) the qubits couple to the waveguide at a rate much smaller than the \gls{CPS}, i.e., $\kappa_{i}\gg \gamma_{i}$,
and (iii) the  \gls{CPS} relaxes to its unique and time-independent steady state $\rho_{\mathrm{CPS}}$, satisfying $\mathcal{L}_{\mathrm{CPS}}\rho_{\mathrm{CPS}}=0$. 
We begin the following analysis by changing into the rotating frame of the drive, as defined in~\cref{Eq:RotFrame}. To correctly split the remaining dynamics into fast and slow parts, we must perform a few preliminary steps. First, we remove the purely coherent driving term from the interaction Lindbladian. This can be achieved by subtracting the steady state expectation values of the coupling operators $\mathcal{L}_{\rm I}$ and reabsorbing them into the qubit Hamiltonian, i.e., 
\begin{align}
\begin{aligned}
    \mathcal{L}_{\rm I}&\to\sum_i \sqrt{\kappa_{i} \gamma_{i}} \Big( [\overline{O}_i\bullet, \sigma_i^{+}] + [\sigma_i^{-},\bullet \overline{O}_i^\dag] \Big), \\
    \mathcal{L}_{\rm Q}&\to \mathcal{L}_{\rm Q}+\sum_i\left[\sqrt{\gamma_{i}}(\varepsilon_i^* \sigma_i^--\text{H.c.}),\bullet  \textbf{}\right],
\end{aligned}
\end{align}
with $\overline{O}_i\equiv O_i-\langle O_i \rangle_{\mathrm{CPS}}$ and $\varepsilon_i=\sqrt{\kappa_{i}}\langle O_i\rangle_{\mathrm{CPS}}=\sqrt{\kappa_{i}}\Tr_{\mathrm{CPS}}[O_i\rho_{\mathrm{CPS}}]$. In the next step, we change into the interaction picture with respect to the target-qubit Hamiltonian $H_\mathrm{Q}=\sum_i\frac{\Delta_{\mathrm{q},i}}{2}\sigma_i^z$, yielding
\begin{align}
\begin{aligned}
\mathcal{L}_{\rm I}\to& \sum_i \sqrt{\kappa_{i} \gamma_{i}} \Big( [\overline{O}_i\bullet, \sigma_i^{+}(t)] + [\sigma_i^{-}(t),\bullet \overline{O}_i^\dag] \Big),  \\
    \mathcal{L}_{\rm Q}\to&  \left[\sum_i \sqrt{\gamma_{i}}(\varepsilon_i \sigma_i^-(t)-\text{H.c.}),\bullet \right]+\sum_i\gamma_{i}\mathcal{D}[\sigma_i^-],
\end{aligned}
\end{align}
with $\sigma_i^-(t)=\sigma_i^-e^{-i\Delta_{\mathrm{q},i}t}$.
This transformation is necessary to guarantee that the evolution generated by $\mathcal{L}_{\rm Q}$ is slow compared to the dynamics under $\mathcal{L}_{\mathrm{CPS}}$. Using standard projector-based adiabatic eliminiation schemes~\cite{breuer_theory_2002,gonzalez-ballestero_tutorial_2024}, we obtain the Born-Markov \gls{ME}
\begin{align}
    &\ddt\rho_{\mathrm{q}}(t)=\mathcal{L}_{\rm Q} \rho_{\rm q}(t) 
    \label{Eq:SimplifiedNZ} \\
    &+\int_{0}^\infty \mathrm{d}\tau \, \Tr_{\rm CPS}\left\{\mathcal{L}_{\rm I}(t)e^{\mathcal{L}_{\mathrm{CPS}} \tau}\mathcal{L}_{\rm I}(t-\tau)\rho_{\mathrm{q}}(t)\otimes \rho_{\mathrm{CPS}}\right\}.\nonumber
\end{align}
We now use the explicit form of $\mathcal{L}_{\rm I}$ and apply the \gls{QRT}~\cite{carmichael_statistical_1999} to evaluate the second-order contribution in~\cref{Eq:SimplifiedNZ},
\begin{align}
\begin{aligned}
        &\Tr_{\mathrm{CPS}}[\mathcal{L}_{\rm I}(t)\int_0^{\infty} \mathrm{d}\tau e^{\mathcal{L}_{\mathrm{CPS}}\tau}\mathcal{L}_{\rm I}(t-\tau)\rho_{\mathrm{q}}(t)\rho_{\mathrm{CPS}}] \\
    &=\sum_{s_1,s_2 =\pm}\sum_{i,j}\sqrt{\gamma_{i}\gamma_{j}}C^{ij}_{s_1s_2}(\Delta_{\mathrm{q},j}) s_1s_2 \times\dots\\
    &\quad\times\left[\left[\bullet,\sigma_j^{-s_1}(t) \right] ,\sigma_i^{-s_2}(t)\right].
\end{aligned}
\end{align}
The coefficients $C_{\pm\pm}^{ij}$ that appear in this expression are the correlation functions defined in~\cref{Eq:BBCoefs}.
Finally, after undoing the rotating frame transformation with respect to $H_\mathrm{Q}$, we obtain the effective \gls{ME} in~\cref{Eq:CPSME}. The compact form in~\cref{Eq:SqueezingLindbladian} is obtained by omitting all nonresonant contributions and by grouping the remaining terms into a Hamiltonian and a dissipative part~\cite{rivas_open_2012}. Furthermore, in~\cref{Eq:SqueezingLindbladian} we have absorbed Lamb shifts of the qubit frequencies into a redefinition of the qubit detunings, \begin{align}
    \Delta_{\mathrm{q},i}\to  \Delta_{\mathrm{q},i}+2\gamma_i \Im{C^{ii}_{-+}(\Delta_{\mathrm{q},i})}.
\end{align}

\section{Beyond the Markovian Regime}\label{Sec:NonMarkovian} 

\begin{figure}[t]
    \centering
    \includegraphics[width=1\linewidth]{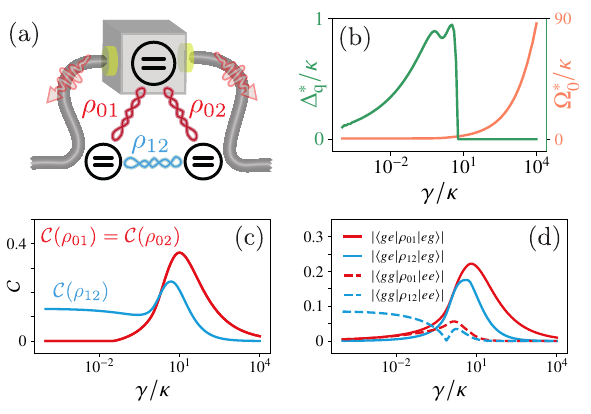}
    \caption{(a) Schematic of the scenario considered in~\cref{Sec:NonMarkovian}. The \gls{CPS} is a \gls{TLS} emitting into the waveguides with two strongly coupled target qubits. In the non-Markovian regime, $\gamma \gtrsim \kappa$, both the target qubits and the TLS of the source can become mutually entangled. (b) Optimal parameters that maximize the concurrence $\mathcal{C}_\mathrm{d}$ for every ratio $\gamma/\kappa$ and for a resonant drive, $\Delta_0=0$, (c) Pairwise concurrence and (d) coherences of the reduced density operators $\rho_{ij}$ between different pairs of qubits $i$ and $j$ [see panel (a)]. All results in these plots have been obtained from exact numerical simulations of the full network \gls{ME} in~\cref{Eq:FullSyst}.
    }
    \label{fig:Mollow_Appendix_1}
\end{figure}

In~\cref{Sec:FiniteBandwidth} we mentioned a scenario where a significant amount of entanglement is distributed between the target qubits for $\kappa \approx \gamma$, which lies outside the Markovian regime considered in the rest of the paper. 
Here, we take a closer look at this particular case by solving the full cascaded \gls{ME} of the network given in~\cref{Eq:FullSyst}, i.e., without adiabatically eliminating the \gls{CPS}. As in~\cref{Sec:DrivenTLS}, we model the latter as a bare driven \gls{TLS} and label it by the index $i=0$. 

In~\cref{fig:Mollow_Appendix_1} we solve the steady state of this system and evaluate the resulting bi-partite entanglement for the different reduced two-qubit states depicted in~\cref{fig:Mollow_Appendix_1}(a) as a function of $\gamma/\kappa$. In these plots, we consider a resonantly driven \gls{TLS}, $\Delta_0=0$, and choose a driving frequency $\Omega_0^*$ and a symmetric detuning $\Delta_{{\rm q},1}=-\Delta_{{\rm q},2}=\Delta_{\rm q}^*$ of the target qubits such that the distributable entanglement, $\mathcal{C}_{\rm d}\equiv\mathcal{C}(\rho_{12})$, is optimized. The corresponding values of $ \Omega^*_0$ and $\Delta_{\rm q}^*$ are plotted in~\cref{fig:Mollow_Appendix_1}(b) for different ratios $\gamma/\kappa$. From the resulting plots of the concurrences in~\cref{fig:Mollow_Appendix_1}(c), we see that the distributable entanglement indeed reaches a value of $\mathcal{C}(\rho_{12})\approx 0.25$, which exceeds the maximally achievable value in the Markovian regime. This maximum is reached for $\gamma/\kappa\approx 4.9$, and a driving strength of $\Omega_0^*/\kappa\approx 1.9$ with the qubits detuned symmetrically by $\Delta_\mathrm{q}^*/\kappa\approx 0.6$. At the same time, near this maximum, the target qubits also become significantly entangled with the source, $\mathcal{C}(\rho_{01})=\mathcal{C}(\rho_{02})\approx 0.32$, indicating that the whole network relaxes into a combined multi-partite entangled steady state.    

To analyze this combined state, we plot in~\cref{fig:Mollow_Appendix_1}(d) a representative set of coherences of the reduced steady states. In the Markov limit, $\gamma/\kappa\ll 1$, there is only one relevant coherence between the states $\ket{gg}_{12}$ and $\ket{ee}_{12}$, no entanglement between the source and the target qubits, and in the optimal configuration, the target qubits are detuned by multiple linewidths. These results are fully consistent with a steady state of the form $\rho\approx \rho_0\otimes |\psi_{12}\rangle\langle \psi_{12}|$, with an entangled qubit state
\begin{equation}
|\psi_{12}\rangle \sim  \ket{gg}_{12}+ \ket{ee}_{12},
\end{equation}
as discussed in the main text. In the non-Markovian regime, $\gamma\gtrsim \kappa$, we find that these coherences are no longer present, but that different superpositions of the form $|eg\rangle \pm|ge\rangle$ become relevant. At the same time, the target qubits become entangled with the source and the optimal detuning drops to $\Delta_{\rm q}^*=0$. The latter implies that the target qubits are now coherently driven by the coherently scattered driving field, i.e., by the terms $\sim \sqrt{\gamma\kappa} \langle \tau^-\rangle_{\rm ss}$ in~\cref{Eq:CPSME}. This situation is similar to the cascaded networks of driven \gls{TLS}s considered in Refs.~\cite{stannigel_driven-dissipative_2012,pichler_quantum_2015,shah_stabilizing_2024,irfan_autonomous_2025}, where superpositions of the form 
\begin{equation}
   |\psi_{01}\rangle \sim |gg\rangle_{01} +(|eg\rangle_{01} - |ge\rangle_{01})
\end{equation}
 appear as dark steady states of the driven-dissipative dynamics. Therefore, we conclude that, with increasing $\gamma/\kappa$, the network transitions from a Markovian regime with squeezing-mediated entanglement to a non-Markovian regime with a singlet-type dark state involving all three subsystems. Of course, this picture is only a rough approximation of the exact steady state, which involves both types of superpositions and is a more complicated mixed state.

\section{Adiabatic elimination of driven \gls{TLS} in QMS regime}\label{Sec:AETLS}
In this appendix, we carry out the adiabatic eliminiation of the driven \gls{TLS} in the \gls{QMS} regime to derive an effective model for the bosonic modes. This derivation follows closely the work by Kustura \etal{}~\cite{kustura_effective_2021} with the main steps outlined here for completeness. 

\subsection{Adiabatic elimination}
We start from the Lindbladian in~\cref{Eq:MollowCavityBB} and first split it up into a part acting solely on the bosonic modes, a part acting only on the \gls{TLS}, and an interaction part, $\mathcal{L}_\mathrm{CPS}=\mathcal{L}_{\rm c}+\mathcal{L}_0+\mathcal{L}_{c0}$, with
\begin{align}
\begin{aligned}
\label{Eq:SplittUp}
    \mathcal{L}_{\rm c}=&\,-i[\Delta_{\mathrm{c},1} a_1^\dag a_1+\Delta_{\mathrm{c},2} a_2^\dag a_2,\bullet] \\
    &\,+\kappa_1 \mathcal{D}[a_1]+\kappa_2 \mathcal{D}[a_2], \\
    \mathcal{L}_{0}=&\,-i\left[\frac{\Delta_{0}}{2} \tau^z+\frac{\Omega_0}{2}(\tau^++\tau^-),\bullet\right]\\
    &\,+\Gamma_0 \mathcal{D}[\tau^-], \\
    \mathcal{L}_{c0}=&\,-i[g \tau^- a_1^\dag+g \tau^- a_2^\dag+\text{H.c.}, \bullet].
\end{aligned}
\end{align}
In order to eliminate the \gls{TLS}, we assume that it relaxes quickly to its steady state $\rho_{0}$ ($\mathcal{L}_{0}\rho_{0}=0$), which is only slightly perturbed by the coupling. We thus require $\Gamma_0 \gg \kappa,g$.

To proceed, we now perform the same transformations as previously used when eliminating the \gls{CPS}, i.e., we first remove the coherent drive on the fast system from the interaction Lindbladian and then transform to the rotating frame with respect to the Hamiltonian part in~\cref{Eq:SplittUp} to get
\begin{align}
\begin{aligned}
    \mathcal{L}_{\rm c}=&\,-i[\varepsilon_1 a_1(t)+\varepsilon_2 a_2(t)+\text{H.c.},\bullet]\\
    &\,+\kappa \mathcal{D}[a_1]+\kappa \mathcal{D}[a_2], \\
    \mathcal{L}_{0}=&\,\Gamma_0 \mathcal{D}[\tau^-], \\
    \mathcal{L}_{c0}=&\,-i[g \overline{\tau}^-(t) a_1^\dag(t)+g \overline{\tau}^- (t)a_2^\dag(t)+\text{H.c.}, \bullet],
\end{aligned}
\end{align}
for $\varepsilon_i=g\langle \tau^+\rangle_{0}=g\Tr_{0}[\tau^+\rho_{0}]$ and $\overline{\tau}^-=\tau^--\langle \tau^-\rangle_{0}$. In this frame, the dominant time-scale is the relaxation of the \gls{TLS}, and we can start the adiabatic elimination from the simplified Born-Markov \gls{ME}~\cite{gonzalez-ballestero_tutorial_2024}. The second-order contribution is given by
\begin{align}
\begin{aligned}
    &\int_{0}^\infty \mathrm{d}\tau \, \Tr_{0}\left\{\mathcal{L}_{c0}(t)e^{\mathcal{L}_{0}\tau}\mathcal{L}_{c0}(t-\tau)\rho_{\rm c}(t)\otimes \rho_{0}\right\} \\
    &=-\int_{0}^\infty \mathrm{d}\tau \, \Tr_{0}\left\{\left[(S^\dag(t)\overline{\tau}^ -+\text{H.c.}),\right. \right. \\
    &\left. \left. e^{\mathcal{L}_{0}\tau}\left[(S^\dag(t-\tau)\overline{\tau}^-+\text{H.c.}),\rho_{\rm c}(t)\otimes \rho_{0}\right] \right]\right\}, 
    \label{Eq:SecondOrderContributionAE}
\end{aligned}
\end{align}
where we defined $S(t)=g a_1 e^{-i\Delta_{\mathrm{c},1} t}+g a_2 e^{-i\Delta_{\mathrm{c},2} t}$. Taking the trace over the fast system, one can rearrange the terms and use the \gls{QRT} to obtain the two-time correlation functions of the transformed \gls{TLS} operators,
\begin{align}
    \Gamma_{s_1,s_2}^{n,m}&=g^2 \int_{0}^{\infty}\mathrm{d} \tau \langle \overline{\tau}^ {s_1}(\tau)\overline{\tau}^ {s_2}(0) \rangle_{0}e^{i s_2\Delta_{\mathrm{c},m} \tau}. 
    \label{Eq:Spectrum}
\end{align}
By rewriting the different terms in~\cref{Eq:SecondOrderContributionAE} as
\begin{align}
\begin{aligned}
 \Tr_{0}\{...\}= \;&-\sum_{m,n}\sum_{s_1,s_2}\Gamma^{n,m}_{-s_1,-s_2}a^{s_1}_n a^{s_2}_m \rho_{0}(t)\\
    &+\sum_{m,n}\sum_{s_1,s_2}(\Gamma^{n,m}_{s_1,s_2})^* a_n^{s_1}\rho_{0}(t)a_m^{s_2}\\
    &-\sum_{m,n}\sum_{s_1,s_2}(\Gamma^{n,m}_{s_1,s_2})^* \rho_{0}(t)a^{s_2}_m a^{s_1}_n\\
    &+\sum_{m,n}\sum_{s_1,s_2}\Gamma^{n,m}_{-s_1,-s_2}a^{s_2}_m \rho_{\mathrm{
    F}}(t)a^{s_1}_n,    
\end{aligned}
\end{align}
where $n,m \in \{1,2\}$, $s_i\in \{+,-\}$ and we defined $a_i^+\equiv a_i^\dag$, $a_i^-\equiv a_i$.

As a last step, we group  the different terms into dissipative and unitary contributions~\cite{rivas_open_2012}. The computations are lengthy, and we only give the final result relating the $\Gamma_{s_1,s_2}^{n,m}$ coefficients to the effective parameters occurring in the quadratic Lindbladian in~\cref{Eq:MollowCavityBB},
\begin{align}
\begin{aligned}
\Gamma_{i,j}=&\;(\Gamma^{i,j}_{-,-})^*+\Gamma^{j,i}_{+,+}, \\
g_{i,j}=&\;-\frac{i}{2}(\Gamma^{j,i}_{+,+}-(\Gamma^{i,j}_{-,-})^*), \\
    \gamma_+^{i,j}=&\;(\Gamma^{j,i}_{+,-})^*+\Gamma^{i,j}_{+,-}, \\
    \gamma_-^{i,j}=&\;(\Gamma^{j,i}_{-,+})^*+\Gamma^{i,j}_{-,+}, \\
    \delta_{i,j}=&\;-\frac{i}{2}(\Gamma^{i,j}_{-,+}+\Gamma^{i,j}_{+,-})+\frac{i}{2}(\Gamma^{j,i}_{-,+}+\Gamma^{j,i}_{+,-})^*.
\end{aligned}
\label{Eq:QuadraticCoefsTLSAE}
\end{align}
The \gls{ME} we obtain in this way is not yet in Lindblad form and thus does not necessarily generate \gls{CPTP} dynamics~\cite{rivas_open_2012,azouit_generic_2017}. Especially in the strong-driving regime, it is necessary to apply a secular approximation by dropping the far-off-resonant interaction terms in order to recover a Lindblad-type generator and thereby guarantee \gls{CPTP} dynamics.

\subsection{Equations of motion and quantum regression theorem} \label{Sec:QRT}
The effective Lindbladian for the cavities in~\cref{Eq:MollowCavityBB} is quadratic, which implies that Gaussian states remain Gaussian under the dynamics. We can therefore fully characterize the state of the two bosonic modes by the set of expectation values $\{\langle a_i\rangle,\langle a_i a_j\rangle,\langle a_i^\dag a_j\rangle\}$. To take advantage of this property, we derive the equations of motion for the different observables using the relation
\begin{align}
    \ddt\langle O_i\rangle(t)=\Tr\left\{\dot{\rho}(t)O_i\right\}.\label{Eq:MomentsEvolution}
\end{align}
The computations are straightforward but tedious and can be performed using symbolic computations in Python~\cite{lim_pybolano_2025}. The first-order moments $\vec{v}=(\langle a_1\rangle,\langle a_1^\dag\rangle,\langle a_2\rangle,\langle a_2^\dag\rangle)^T$ evolve according to
\begin{align}
    \dot{\vec{v}}(t)=\mathcal{M} \vec{v}(t)+\vec{f}. \label{eq:EQMMoments}
\end{align}
To compute the emission spectrum of the cavities 
\begin{align}
    \Gamma_{O_1,O_2}(\omega)=\int_0^\infty \mathrm{d}t \langle \overline{O}_1(t)O_2(0)\rangle_{\mathrm{CPS}} e^{-i\omega t},\label{Eq:CorrfuncLaplace}
\end{align}
for $O_i \in \{a_1,a_1^\dag,a_2,a_2^\dag\}$ and $\langle  O \rangle_{\mathrm{CPS}}=\Tr[O\rho_{\mathrm{CPS}}]$, we invoke the \gls{QRT}~\cite{carmichael_statistical_1999}, which states that
\begin{align}
    \langle \vec{\overline{v}}(t)O_2(0) \rangle_{\mathrm{CPS}}=e^{\mathcal{M} t}\langle \vec{\overline{v}}O_2 \rangle_{\mathrm{CPS}}^\mathrm{ss},
\end{align}
where $\langle \vec{\overline{v}}O_2 \rangle_{\mathrm{CPS}}^\mathrm{ss}$ can be computed from the steady state of the coupled differential equations given in~\cref{Eq:MomentsEvolution}. Inserting this result into~\cref{Eq:CorrfuncLaplace} and using the fact that the eigenvalues of $\mathcal{M}$ have a negative real part for a stable system, we obtain
\begin{align}
    \Gamma_{\vec{v},O_2}(\omega)&=\frac{1}{i\omega-\mathcal{M}}\langle \vec{\overline{v}}O_2 \rangle_{\mathrm{CPS}}^\mathrm{ss}.
\end{align}

\bibliographystyle{apsrev4-2}
\bibliography{refs.bib}
\end{document}